\documentclass[10pt]{revtex4}
\usepackage{amssymb}
\usepackage{latexsym}
\usepackage{epsfig}

\begin{document}

\title{ Birth of the GUP and its effect on the entropy of the Universe in Lie-$N$-algebra  }

\author{Alireza Sepehri $^{1,2}$ \footnote{alireza.sepehri@uk.ac.ir}}
 \affiliation{ $^{1}$Faculty of
Physics, Shahid Bahonar University, P.O. Box 76175, Kerman,
Iran.\\$^{2}$ Research Institute for Astronomy and Astrophysics of
Maragha (RIAAM), P.O. Box 55134-441, Maragha, Iran. }

\author{Anirudh Pradhan $^{3}$\footnote{pradhan.anirudh@gmail.com}}
\affiliation{$^{3}$Department of Mathematics, Institute of Applied Sciences and Humanities, G L A
 University, Mathura-281 406, Uttar Pradesh, India.}

\author{Richard Pincak $^{4,5}$ \footnote{pincak@saske.sk}}
 \affiliation{  $^{4}$ Institute of Experimental Physics,  Slovak Academy of Sciences,
Watsonova 47,  043 53 Kosice, Slovak Republic.  \\ $^{5}$
Bogoliubov Laboratory of Theoretical Physics,  Joint
Institute for Nuclear Research, 141980 Dubna,  Moscow region,  Russia.
 }

\author{Farook Rahaman $^{6}$ \footnote{rahaman@associates.iucaa.in}}
 \affiliation{$^{6}$ Department of Mathematics, Jadavpur University, Kolkata 700 032, West Bengal,
India.}

\author{A. Beesham $^{7}$\footnote{beeshama@unizulu.ac.za}} 
\affiliation{ $^{4}$ Department of Mathematical Sciences, University of 
Zululand, Kwa-Dlangezwa 3886, South Africa.}

\author{Tooraj Ghaffary $^{8}$\footnote{ghaffary@iaushiraz.ac.ir}} 
\affiliation{ $^{8}$ Department of Sciences, Shiraz Branch, Islamic Azad University, Shiraz, Iran}

\begin{abstract}
 In this paper, the origin of the generalized uncertainty principle (GUP) in an $M$-dimensional theory  with 
 Lie-$N$-algebra is considered. This theory which we name GLNA(Generalized Lie-$N$-Algebra)-theory 
 can be reduced to $M$-theory with $M=11$ and $N=3$. In this theory, at the beginning, two  energies with 
 positive and negative signs are created from nothing and produce two types of branes with opposite quantum 
 numbers and different numbers of timing dimensions. Coincidence with the birth of these branes, various derivatives 
 of bosonic fields emerge in the action of the system which produce the $r$ GUP for bosons. These branes interact 
 with each other, compact and various derivatives of spinor fields appear in the action of the system which 
 leads to the creation of the GUP for fermions. The previous  predicted entropy of branes in the GUP is corrected 
 as due to the emergence of higher orders of derivatives and different number of timing dimensions.
\vspace{5mm}


\vspace{5mm}


 \end{abstract}
 \date{\today}

\maketitle
\section{Introduction}
 An interesting result of respective hypotheses of quantum gravity, $M$-theory and field theory  is
the creation of a minimum measurable length. This leads to a modification of the Heisenberg uncertainty
principle (HUP) or equivalently, modified commutation relations between position coordinates and momenta, 
which is well known as the generalized uncertainty principle (GUP)\cite{g1,g2,g3,g4,g5,g6,g7}. There are 
different versions of the GUP. One version and the comparable modified commutators between position and 
momentum coordinates contain the quadratic terms, and are given by \cite{g8,g9}:

\begin{eqnarray}
 && \Delta x_{i}\Delta p_{i}\geq \frac{\hbar}{2}[1+\beta\Big((\Delta p)^{2}+<p>^{2}\Big)+\nonumber\\&& 
 2\beta\Big((\Delta p_{i})^{2}+<p_{i}>^{2}\Big)],\quad i=1,2,3 \nonumber\\&& [x_{i},p_{i}]=i\hbar(\delta_{ij}+\beta \delta_{ij}p^{2}+
 2\beta p_{i}p_{j})\label{f1}
\end{eqnarray}
where $p^{2}=\Sigma_{j=1}^{3}p_{i}p_{j}$, $\beta=\frac{\beta_{0}}{M_{P}c^{2}}$ and $M_{P}$ is the Planck mass. On the other hand,  
another version of the GUP has been proposed which is uniform with the above as well as with Doubly special 
relativity (DSR) theories, and includes linear terms in additional to the quadratic terms \cite{g10,g11,g12,g13,g14}:

\begin{eqnarray}
 && \Delta x_{i}\Delta p_{i}\geq \frac{\hbar}{2}[1+\Big(\frac{\alpha}{\sqrt{<p^{2}>}}+4\alpha^{2}\Big)\Delta p^{2}+4\alpha^{2}<p>^{2}-
 2\alpha \sqrt{<p^{2}>}],\quad i=1,2,3 \nonumber\\&& [x_{i},p_{i}]=i\hbar\Big(\delta_{ij}-\alpha \Big(\delta_{ij}p+\frac{p_{i}p_{j}}{p^{2}}\Big)+
 \alpha^{2} \Big(\delta_{ij}p^{2}+3 p_{i}p_{j}\Big)\Big)\label{f2}
\end{eqnarray}
where $\alpha=\frac{\alpha_{0}}{M_{P}c}$. 

Now, the question that arises is  what is the reason for  the emergence of terms with 
different orders in the GUP. We answer this question in an $M$-dimensional theory with Lie-$N$-algebra. We name it \textbf{GLNA-theory}.
In this theory, firstly,  two types of energies with opposite signs are produced, excited and create two types of branes with 
differing numbers of timing dimensions and quantum numbers. During the formation of branes, various bosonic fields emerge such 
that the order of  derivatives in their time-spatial wave equations changes from zero to higher numbers  
corresponding to the number of dimensions of branes ($p$), the universe ($M$) and the algebra ($N$). This leads to the 
appearance of  terms with  different orders of momenta in energy-momentum space. On the other hand, by compacting branes, 
fermions emerge and their wave equations contain various orders of derivatives. Thus, different versions of the  GUP are 
produced in which their shapes depend on the number of dimensions of branes and the algebra. Each of these versions produces 
a special entropy for which the order of terms in it grows with the increase in the number of dimensions of the brane.

In section \ref{o1}, we consider  the process of the emergence of GUP during the formation of branes in the Lie-$N$-algebra. 
In section \ref{o2}, we have obtained the corrected entropy by using the exact form of the GUP for a $p$-dimensional brane. The summary 
and conclusion are given in the last section.
The units used throughout the paper are: $\hbar=c=8\pi G=1$.

\section{Emergence of GUP in Lie-N-algebra  }\label{o1}

In this section, first, we obtain the action of $p$ dimensional branes in a universe with M-dimensions and Lie-$N$-algebra. Then, 
by extracting the wave equation of motion, we show that various derivatives appear for which their orders change from zero 
to the dimensions of the branes. After that, by replacing these derivatives with momenta, we obtain exact forms of the GUP for each 
brane in the universe. We observe that the number of terms and their orders depend on the dimension of the branes and the algebra.

Firstly, we should introduce the general form of actions in Lie-$N$-algebra. Previously, it has been shown that all 
$Dp$-branes in string theories are built from $D0$-branes which follow from Lie-two-algebras with two dimensional 
brackets \cite{h2,h3,h4,h5,h6,h7,h8,h9,h10,h11,h12}. Also, all $Mp$-branes in $M$-theory are constructed from $M0$-branes 
which arise from a Lie-three-algebra with three dimensional brackets \cite{h13,h14,h15,h16}. Now, by using the Lie-$N$-algebra and 
generalizing  dimensions to $M$, we construct a new theory which includes all the properties of string theory and $M$-theory 
and resolve the puzzles in them. To show this, we begin of the action for the Dp-brane \cite{h9,h10,h11,h12}:

\begin{eqnarray}
 &&S=-\frac{T_{Dp}}{2}\int d^{p+1}x \sum_{n=1}^{p}\beta_{n} \chi^{\mu_{0}}_{[\mu_{0}}\chi^{\mu_{1}}_{\mu_{1}}...\chi^{\mu_{n}}_{\mu_{n}]}\label{s4}
\end{eqnarray}

\begin{eqnarray}
\chi^{\mu}_{\nu}\equiv  \delta ^{\mu}_{\nu} STr \Bigg(-det(P_{ab}[E_{mn}
E_{mi}(Q^{-1}+\delta)^{ij}E_{jn}]+ \lambda F_{ab})det(Q^{i}_{j})\Bigg)^{1/2}~~
\label{r1}
\end{eqnarray}

where

\begin{eqnarray}
   E_{mn} = G_{mn} + B_{mn}, \qquad  Q^{i}_{j} = \delta^{i}_{j} + i\lambda[X^{j},X^{k}]E_{kj} \label{r2}
\end{eqnarray}
 $\lambda=2\pi l_{s}^{2}$ and $G_{ab}=\eta_{ab}+\partial_{a}X^{i}\partial_{b}X^{i}$ and $X^{i}$ are  scalar strings that link  
 to branes. In this equation, $a,b=0,1,...,p$
are the world-volume indices of the $Dp$-branes, $i,j,k = p+1,...,9$
are the indices of the transverse space, and m,n are related to
ten-dimensional spacetime indices. Also, $T_{Dp}=\frac{1}{g_{s}(2\pi)^{p}l_{s}^{p+1}}$ refers the tension of the 
$Dp$-brane, $l_{s}$ is the string length and $g_{s}$ refers to the string
coupling. Now, we can show that the action of the $Dp$-brane can be constructed by summing 
over actions of $D0$-branes. To this end, we make use of the rules below \cite{h2,h4,h9,h10,h11,h12}:

\begin{eqnarray}
&& \Sigma_{a=0}^{p}\Sigma_{m=0}^{9}\rightarrow \frac{1}{(2\pi l_{s})^{p}}\int d^{p+1}\sigma \Sigma_{m=p+1}^{9}\Sigma_{a=0}^{p} 
\qquad \lambda = 2\pi l_{s}^{2}\nonumber \\
&& i,j=p+1,..,9\qquad a,b=0,1,...p\qquad m,n=0,1,..,9 \nonumber \\
&& i,j\rightarrow a,b \Rightarrow [X^{a},X^{i}]=i \lambda
\partial_{a}X^{i}\qquad  [X^{a},X^{b}]=\frac{ i \lambda F^{ab}}{2} \nonumber \\
&& \frac{1}{Q}\rightarrow \sum_{n=1}^{p}
\frac{1}{Q}(\partial_{a}X^{i}\partial_{b}X^{i}+\frac{\lambda^{2}}{4} (F^{ab})^{2})^{n}\nonumber \\
&& det(Q^{i}_{j})\rightarrow det(Q^{i}_{j})\prod_{n=1}^{p}
det(\partial_{a_{n}}X^{i}\partial_{b_{n}}X^{i}+\frac{\lambda^{2}}{4} (F^{a_{n}b_{n}})^{2}) \label{r3}
\end{eqnarray}

Using the above laws, we can show that the action of Dp-branes can be written in terms of two dimensional 
brackets \cite{h2,h4,h9,h10,h11,h12}:

\begin{eqnarray}
&& S_{Dp} = -(T_{D0})^{p} \int dt \sum_{n=1}^{p}\beta_{n}\Big(
\delta^{a_{1},a_{2}...a_{n}}_{b_{1}b_{2}....b_{n}}L^{b_{1}}_{a_{1}}...L^{b_{n}}_{a_{n}}\Big)^{1/2}\nonumber\\&&
(L)^{b}_{a}=Tr\Big( \Sigma_{a,b=0}^{p}\Sigma_{j=p+1}^{9}(
[X^{a},X^{j}][X_{b},X_{j}]+[X^{a},X^{b}][X_{b},X_{a}]+[X^{i},X^{j}][X_{i},X_{j}])\Big) \label{s7}
\end{eqnarray}
where we have made use of the antisymmetric properties for $\delta$. At this stage, we can show that the  above action is 
built by summing over the actions of $D0$-branes by using the definition below for  the $D0$-brane \cite{h2,h3,h4,h5,h6,h7,h8,h9,h10,h11,h12}:

\begin{eqnarray}
&& S_{D0} = -T_{D0} \int dt Tr( \Sigma_{m=0}^{9}
[X^{m},X^{n}]^{2}) \label{s8}
\end{eqnarray}

Replacing the two dimensional brackets in the Lie-two-algebra by three dimensional brackets in the Lie-three-algebra, we obtain the 
action of the $M0$-branes\cite{h2,h3,h4,h5,h6,h7,h8,h9,h13,h14,h15,h16}:

\begin{eqnarray}
S_{M0} = T_{M0}\int dt Tr( \Sigma_{M,N,L=0}^{10}
\langle[X^{M},X^{N},X^{L}],[X^{M},X^{N},X^{L}]\rangle) \label{s9}
\end{eqnarray}
where $X^{M}=X^{M}_{\alpha}T^{\alpha}$ and

\begin{eqnarray}
 &&[T^{\alpha}, T^{\beta}, T^{\gamma}]= f^{\alpha \beta \gamma}_{\eta}T^{\eta} \nonumber \\&&\langle T^{\alpha}, 
 T^{\beta} \rangle = h^{\alpha\beta} \nonumber \\&& [X^{M},X^{N},X^{L}]=[X^{M}_{\alpha}T^{\alpha},X^{N}_{\beta}T^{\beta},X^{L}_{\gamma}T^{\gamma}]
 \nonumber \\&&\langle X^{M},X^{M}\rangle = X^{M}_{\alpha}X^{M}_{\beta}\langle T^{\alpha}, T^{\beta} \rangle
\label{s10}
\end{eqnarray}
where  $X^{M}$(i=1,3,...10) are  scalar strings which are linked to the M0-brane. By substituting $N$-dimensional brackets 
instead of three dimensional brackets in the action of (\ref{s8}), we obtain the action of the $G0$-brane in \textbf{GLNA-theory} as:

 \begin{eqnarray}
 S_{G0} = T_{G0}\int dt Tr( \Sigma_{L_{1}=L_{2}..L_{N}=0}^{M}
 \langle[X^{L_{1}},X^{L_{2}},...X^{L_{N}}],[X^{L_{1}},X^{L_{2}},...X^{L_{N}}]\rangle) \label{P1}
 \end{eqnarray}
 where $X^{M}=X^{M}_{\alpha}T^{\alpha}$ and

 \begin{eqnarray}
  &&[T^{\alpha_{1}}, T^{\alpha_{2}}..T^{\alpha_{N}}]= f^{\alpha_{1}..\alpha_{N}}_{\alpha_{L}}T^{L} \nonumber \\&&\langle T^{\alpha}, 
  T^{\beta} \rangle = h^{\alpha\beta} \nonumber \\&& [X^{L_{1}},X^{L_{2}},...X^{L_{N}}]=[X^{L_{1}}_{\alpha_{1}}T^{\alpha_{1}},X^{L_{2}}_{\alpha_{2}}
  T^{\alpha_{2}},...X^{L_{N}}_{\alpha_{N}}T^{\alpha_{N}}]\nonumber \\&&\langle X^{M},X^{M}\rangle = X^{M}_{\alpha}X^{M}_{\beta}\langle T^{\alpha}, 
  T^{\beta} \rangle
 \label{P2}
 \end{eqnarray}

  The above action reduces to the action of $M0$-branes by putting $N=3$ and $M=10$ and also to the action of $D0$-branes for 
  $N=2$ and $M=9$. Now, the question that arises is what is the origin of this brane in \textbf{GLNA-theory}?  To answer 
  this question, we suppose  that first, two energies with opposite signs are produced such that the sum over them is zero. 
  Then, these energies  create $2M$ degrees of freedom which each two of them cause  to creation of new dimension. After that,
  $M-N$ of degrees of freedom are hidden by compacting half of M-N dimensions on a circle to create Lie-$N$-algebra. During 
  this process, the behaviour of some dimensions changes and they form times. Also, for second energy, the properties 
  of more dimensions change which lead to the emergence of more time coordinates. Thus, the physics of branes is completely 
  different from anti-branes.

         Let us to begin with two oscillating energies which  are created from nothing and expands in the  $M^{th}$ dimension. We obtain:

     \begin{eqnarray}
     && E \equiv 0\equiv E_{1}+E_{2}\equiv 0\equiv N_{1}+N_{2}\equiv k((X^{M})^{2}-  (X^{M})^{2}) = k \int d^{2}x (\frac{\partial}
     {\partial x})^{2}((X^{M})^{2}-  (X^{M})^{2})
     \label{t1}
     \end{eqnarray}
         where $N_{1/2}$ are the number of degrees of freedom for first and second energies. These energies are oscillating, 
         excited and produce $M$ dimensions with $2M$ degrees of freedom. This can be shown clearly  by rewriting equation (\ref{t1}) as 
         follows:

      \begin{eqnarray}
      && E \equiv 0\equiv k \int d^{2M}x \varepsilon^{i_{1}i_{2}...i_{M}}\varepsilon^{i_{1}i_{2}...i_{M}} (\frac{\partial}{\partial x_{i_{1}}}
      \frac{\partial}{\partial x_{i_{2}}}..\frac{\partial}{\partial x_{i_{M-1}}})^{2}(X^{M})^{2}-\nonumber \\ &&  k \int d^{2M}x 
      \varepsilon^{i_{1}i_{2}...i_{M}}\varepsilon^{i_{1}i_{2}...i_{M}} (\frac{\partial}{\partial x_{i_{1}}}\frac{\partial}
      {\partial x_{i_{2}}}..\frac{\partial}{\partial x_{i_{M-1}}})^{2}(X^{M})^{2}
      \label{t2}
      \end{eqnarray}
    where, we apply the definition  $\varepsilon^{i_{1}i_{2}...i_{M}}\varepsilon^{i_{1}i_{2}...i_{M}}=-1$. We can substitute some brackets 
    instead of derivatives and write \cite{h2,h3,h4,h5,h6,h13,h14,h15,h16}:

    \begin{eqnarray}
    && \frac{\partial}{\partial x_{i_{1}}}X^{M}=[ X^{i_{1}},X^{14}] \nonumber \\ && \frac{\partial}{\partial x_{i_{1}}}\frac{\partial}
    {\partial x_{i_{2}}}X^{M}=[ X^{i_{1}},X^{i_{2}},X^{M}]\nonumber \\ && (\frac{\partial}{\partial x_{i_{1}}}\frac{\partial}
    {\partial x_{i_{2}}}..\frac{\partial}{\partial x_{i_{M-1}}})(X^{M})=[ X^{i_{1}},X^{i_{2}},...,X^{i_{M-1}},X^{M}]\nonumber \\ &&
    \varepsilon^{i_{1}i_{2}...i_{M}}\varepsilon^{i_{1}i_{2}...i_{M}} (\frac{\partial}{\partial x_{i_{1}}}\frac{\partial}
    {\partial x_{i_{2}}}..\frac{\partial}{\partial x_{i_{M-1}}})^{2}(X^{M})^{2}=\nonumber \\ && \varepsilon^{i_{1}i_{2}...i_{M}}
    \varepsilon^{i'_{1}i'_{2}...i'_{M}} [(\frac{\partial}{\partial x_{i_{1}}}\frac{\partial}{\partial x_{i_{2}}}..\frac{\partial}
    {\partial x_{i_{M-1}}})(X^{M})][(\frac{\partial}{\partial x_{i'_{1}}}\frac{\partial}{\partial x_{i'_{2}}}..\frac{\partial}
    {\partial x_{i'_{M-1}}})(X^{M})]=\nonumber \\ && \langle [ X_{i_{1}},X_{i_{2}},...,X_{i_{M}}],[ X_{i_{1}},X_{i_{2}},
    ...,X_{i_{M}}] \rangle
    \label{t3}
   \end{eqnarray}

   Applying  the relations of Eq. (\ref{t3}) in Eq. (\ref{t2}), we get:

    \begin{eqnarray}
    && E \equiv 0\equiv E_{1}+E_{2}\equiv \nonumber \\ &&  E_{1}=  k \int d^{2M}x 
    \langle [ X_{i_{1}},X_{i_{2}},...,X_{i_{M}}],[ X_{i_{1}},X_{i_{2}},...,X_{i_{M}}] \rangle \nonumber \\ &&   
    E_{2}=-k \int d^{2M}x \langle [ X_{i_{1}},X_{i_{2}},...,X_{i_{M}}],[ X_{i_{1}},X_{i_{2}},...,X_{i_{M}}] \rangle
     \label{t4}
   \end{eqnarray}

  The shape of the above energies is similar to the shape of the action of $Gp$-branes, however  their algebra has an $M$ dimensional bracket, 
  while the $G0$-brane has an $N$ dimensional bracket and for producing this algebra, we should remove $M-N$  degrees of freedom by compactification.  
  To achieve this aim, we apply the mechanism in  \cite{h13,h14,h15,h16} and replace $X_{i_{n=1,3,5..M-N}}$ by $i T^{i_{n}}\frac{R}{l_{P}^{1/2}}$, 
  where $l_{P}$ is the Planck length. We derive the action below for the first energy:

   \begin{eqnarray}
   && E_{1}\equiv k \int d^{2M}x \langle [ X_{i_{1}},X_{i_{2}},...,X_{i_{M}}],
   [X_{i_{1}},X_{i_{2}},...,X_{i_{M}}] \rangle=
   \nonumber\\&& k \int d^{2M}x  \varepsilon^{i_{1}i_{2}...i_{M}}\varepsilon^{i'_{1}i'_{2}...i'_{M}}X_{i_{1}}X_{i_{2}}...X_{i_{M}}
   X_{i'_{1}}X_{i'_{2}}...X_{i'_{M}} = \nonumber \\ && (i)^{2(M-N)}k \int d^{N}x (\frac{R^{M-N}}{l_{P}^{(M-N)/2}}) \varepsilon^{j_{1}...j_{N}}
   \varepsilon^{j'_{1}...j'_{N}}X_{j_{1}}...X_{j_{N}}X_{j'_{1}}...
   X_{j'_{N}}=\nonumber\\ && (i)^{2(M-N)}k \int d^{N}x (\frac{R^{M-N}}{l_{P}^{(M-N)/2}})
   \langle [ X_{j_{1}},X_{j_{2}},...,X_{j_{N}}],[ X_{j_{1}},X_{j_{2}},...,X_{j_{N}}]\rangle=\nonumber\\ && 
   k \int d^{N}x (\frac{R^{M-N}}{l_{P}^{(M-N)/2}}) \langle [i X_{j_{1}},iX_{j_{2}},...,iX_{j_{M-N}}..,X_{j_{N}}],
   [ iX_{j_{1}},X_{j_{2}},..,...,iX_{j_{M-N}}.,X_{j_{N}}]\rangle
   \label{t5}
   \end{eqnarray}
   where we have defined $ \varepsilon^{1i_{2}...i_{M}}\varepsilon^{1i'_{2}...i'_{M}}=(-i)^{N-M}\varepsilon^{j_{1}...j_{N}}
   \varepsilon^{j'_{1}...j'_{N}}$. Some scalar strings take  one extra (i) and  expand in the time directions. Obviously, in 
   \textbf{GLNA-theory}, there exists $M-N$ time coordinates where $M$ is the dimension of the universe and $N$ is the dimension of the algebra. The 
   reason that we only observe one dimension is  our living in a four dimensional universe and observing the  three dimensional brackets 
   of $M$-theory. For this reason, for us, $M=4$ and $n=3$ and thus we have only one time dimension. For the  second energy, for which its sign 
   is opposite to that of first energy, we have some extra time dimensions:

    \begin{eqnarray}
    && E_{2}=-E_{1}=(i)^{2}E_{1}= \nonumber\\ && k \int d^{N}x (\frac{R^{M-N}}{l_{P}^{(M-N)/2}}) 
    \langle [i X_{j_{1}},iX_{j_{2}},...,iX_{j_{M-N+1}}..,X_{j_{N}}],[ iX_{j_{1}},X_{j_{2}},..,...,iX_{j_{M-N+1}}.,X_{j_{N}}]\rangle  \label{t6}
    \end{eqnarray}

  Thus, properties of anti-branes  which are created by this energy, are different and we have more time dimensions. For example, in 
  one four dimensional anti-universe, there exists  two time coordinates and all things are changed. In our universe, the length of an object 
  can be defined by $l^{2}=-t^{2}+x_{1}^{2}+x_{2}^{2}+x_{3}^{2}$ where $t$ is time and $x_{i}$ are coordinates of space, while, in an anti-universe, 
  length is obtained by $\tilde{l}^{2}=-t_{1}^{2}-x_{1}^{2}+x_{2}^{2}+x_{3}^{2}$. Also, energy and momentum, which is related to the mass with 
  the equation for our universe ($m^{2}=-E^{2}+P_{1}^{2}+P_{2}^{2}+P_{3}^{2}$), has this relation ($m^{2}=-E^{2}-P_{1}^{2}+P_{2}^{2}+P_{3}^{2}$) 
  for an anti-universe.

  Until now, we have considered only symmetrical compactification. However, maybe during this process, the symmetry of the system is broken and only the 
  upper or the lower part of one dimension is compactified, and fermions  emerge ($X \rightarrow \psi^{U}\psi^{L}$ ) \cite{h3}. To include 
  non-symmetrical compactification, we use the mechanism in \cite{h4}, and compactify the $M^{th}$  dimension of the branes on a circle with radius $R$ by 
  choosing  $<X^{M}>=i\frac{R}{l_{p}^{1/2}}T^{M}$ for  bosons and $<\psi^{L,M}>=i\frac{R^{1/2}}{l_{p}^{1/4}}T^{L,M}$ for fermions in the action 
  of (\ref{t5}). We get:

  \begin{eqnarray}
  && E_{1}\equiv  k \int d^{N}x (\frac{R^{M-N}}{l_{P}^{(M-N)/2}})\Big( \langle [i X_{j_{1}},iX_{j_{2}},...,iX_{j_{M-N}}..,X_{j_{N}}],
  [ iX_{j_{1}},X_{j_{2}},..,...,iX_{j_{M-N}}.,X_{j_{N}}]\rangle   \nonumber \\ &&
  - i(\frac{R^{M-N}}{l_{P}^{(M-N)/2}})\langle [i X_{j_{1}},iX_{j_{2}},...,iX_{j_{M-N}}..T_{j_{m'}}.,
  \psi_{R,j_{N}}], [i X_{j_{1}},iX_{j_{2}},...,iX_{j_{M-N}}..T_{j_{m'}}.,\psi_{R,j_{N}}]\rangle ) )\Big)
   \label{P6}
  \end{eqnarray}

  By choosing $\gamma_{j_{m}}=T_{j_{m}}\frac{R^{2}}{l_{p}}$, where the $\gamma_{j_{m}}$'s are the Pauli matrices in $M$ dimensions, we 
  obtain the  action of the initial energy as follows:

 \begin{eqnarray}
 && E_{1}\equiv  k \int d^{N}x  (\frac{R^{M-N}}{l_{P}^{(M-N)/2}})\Big(\langle [i X_{j_{1}},iX_{j_{2}},...,iX_{j_{M-N}}..,X_{j_{N}}],
 [ iX_{j_{1}},X_{j_{2}},..,...,iX_{j_{M-N}}.,X_{j_{N}}]\rangle  \nonumber \\ &&
 -  i\langle [i X_{j_{1}},iX_{j_{2}},...,iX_{j_{M-N}}..\gamma_{j_{m'}}.,\psi_{R,j_{N}}], [i X_{j_{1}},iX_{j_{2}},...,iX_{j_{M-N}}..\gamma_{j_{m'}}.,
 \psi_{R,j_{N}}]\rangle )\Big)
 \label{P7}
 \end{eqnarray}
  This action includes  both fermionic and bosonic degrees of freedom and supersymmetry emerges. If we assume that all scalars depend 
  only on one time coordinate and ($R=l_{P}^{1/2}$), we achieve the action of $G0$-branes:

 \begin{eqnarray}
&& S_{G0}\equiv  k V_{N-1} \int dt \Big(\langle [i X_{j_{1}},iX_{j_{2}},...,iX_{j_{M-N}}..,X_{j_{N}}],
[ iX_{j_{1}},X_{j_{2}},..,...,iX_{j_{M-N}}.,X_{j_{N}}]\rangle   \nonumber \\ &&
  -  i\langle [i X_{j_{1}},iX_{j_{2}},...,iX_{j_{M-N}}..\gamma_{j_{m'}}.,\psi_{R,j_{N}}], [i X_{j_{1}},iX_{j_{2}},...,iX_{j_{M-N}}..
  \gamma_{j_{m'}}.,\psi_{R,j_{N}}]\rangle )\Big)
  \label{PP7}
\end{eqnarray}
 where $V_{N-1}$ is the volume of space which is formed by the remaining coordinates. Also, by adding one negative sign and using 
 equation (\ref{t6}), we obtain the action of the anti-$G0$-brane:

    \begin{eqnarray}
   && S_{Anti-G0}\equiv  k V_{N-1} \int dt  \Big(\langle [i X_{j_{1}},iX_{j_{2}},...,iX_{j_{M-N+1}}..,X_{j_{N}}],
   [ iX_{j_{1}},X_{j_{2}},..,...,iX_{j_{M-N+1}}.,X_{j_{N}}]\rangle   \nonumber \\ &&
    -  i\langle [i X_{j_{1}},iX_{j_{2}},...,iX_{j_{M-N+1}}..\gamma_{j_{m'}}.,\psi_{R,j_{N}}], [i X_{j_{1}},iX_{j_{2}},...,iX_{j_{M-N+1}}..
    \gamma_{j_{m'}}.,\psi_{R,j_{N}}]\rangle )\Big)
    \label{PPP7}
   \end{eqnarray}
  These actions for the $G0$-brane and the $G0$-anti-brane contain both bosonic and fermionic fields which are created  due to symmetrical 
  or non-symmetrical compactification of dimensions. Also, time which is a puzzle in cosmology, is produced by compactification. Thus, we conclude 
  that all fields have the same origin and  begin from nothing. Then, by different compactification, different shapes of matter emerge.

  By substituting N-dimensional brackets in equations (\ref{PP7}) and (\ref{PPP7}) instead of two dimensional brackets, and 
  increasing dimensions from $10$ to $M$ in action (\ref{s7}), we can obtain the action of the  $Gp$-brane and anti-$Gp$-brane:

  \begin{eqnarray}
  &&S_{Gp} = -(T_{G0})^{p} \int dt \sum_{n=1}^{p}\beta_{n}\Big(
 \delta^{a_{1},a_{2}...a_{n}}_{b_{1}b_{2}....b_{n}}L^{b_{1}}_{a_{1}}...L^{b_{n}}_{a_{n}}\Big)^{1/2}\nonumber\\&&
  (L)^{a_{n}}_{b_{n}}= \delta^{a_{n}}_{b_{n}}Tr\Big( \Sigma_{L=0}^{N} \Sigma_{H=0}^{N-L}\Sigma_{a_{1}..a_{L}=0}^{p}\Sigma_{j_{1}..j_{H}=p+1}^{M}(\nonumber \\
  && i^{2(p-N)}\langle[X^{j_{1}},..X^{j_{H-1}},X^{a_{1}},..X^{a_{L}},X^{j_{H}}],\langle[X^{j_{1}},..X^{j_{H-1}},X^{a_{1}},..X^{a_{L}},X^{j_{H}}]\rangle) + 
  \nonumber \\ && i^{2(p-N)}\Sigma_{L=0}^{N} \Sigma_{H=0}^{N-L}\Sigma_{a_{1}..a_{L}=0}^{p}\Sigma_{j_{1}..j_{H}=p+1}^{M}
  (\langle[X^{j_{1}},..X^{j_{H}},X^{a_{1}},..X^{a_{L}}],[X^{j_{1}},..X^{j_{H}},X^{a_{1}},..X^{a_{L}}]\rangle)- \nonumber \\
   &&  i^{2(p-N)+1}\langle[\gamma^{j_{1}},..X^{j_{H-1}},X^{a_{1}},..X^{a_{L}},\psi^{R,j_{H}}],\langle[X^{j_{1}},..X^{j_{H-1}},X^{a_{1}},..X^{a_{L}},
   \psi^{R,j_{H}}]\rangle) - \nonumber \\
   && i^{2(p-N)+1}\Sigma_{L=0}^{N} \Sigma_{H=0}^{N-L}\Sigma_{a_{1}..a_{L}=0}^{p}\Sigma_{j_{1}..j_{H}=p+1}^{M}
  (\langle[\gamma^{j_{1}},..\psi^{R,j_{H}},X^{a_{1}},..X^{a_{L}}],[X^{j_{1}},..\psi^{R,j_{H}},X^{a_{1}},..X^{a_{L}}]\rangle)\Big) \label{P88}
   \end{eqnarray}

  \begin{eqnarray}
  &&S_{Anti-Gp} = -(T_{Anti-G0})^{p} \int dt \sum_{n=1}^{p}\beta_{n}\Big(\delta^{a_{1},a_{2}...a_{n}}_{b_{1}b_{2}....b_{n}}
  L^{b_{1}}_{a_{1}}...L^{b_{n}}_{a_{n}}\Big)^{1/2}\nonumber\\&&
  (L)^{a_{n}}_{b_{n}}= \delta^{a_{n}}_{b_{n}}Tr\Big( \Sigma_{L=0}^{N} \Sigma_{H=0}^{N-L}\Sigma_{a_{1}..a_{L}=0}^{p}
  \Sigma_{j_{1}..j_{H}=p+1}^{M}(\nonumber \\
   && i^{2(p-N+1)}\langle[X^{j_{1}},..X^{j_{H-1}},X^{a_{1}},..X^{a_{L}},X^{j_{H}}],\langle[X^{j_{1}},..X^{j_{H-1}},X^{a_{1}},..X^{a_{L}},X^{j_{H}}]
   \rangle) + \nonumber \\
  && i^{2(p-N+1)}\Sigma_{L=0}^{N} \Sigma_{H=0}^{N-L}\Sigma_{a_{1}..a_{L}=0}^{p}\Sigma_{j_{1}..j_{H}=p+1}^{M}
  (\langle[X^{j_{1}},..X^{j_{H}},X^{a_{1}},..X^{a_{L}}],[X^{j_{1}},..X^{j_{H}},X^{a_{1}},..X^{a_{L}}]\rangle)- \nonumber \\
  &&  i^{2(p-N+1)+1}\langle[\gamma^{j_{1}},..X^{j_{H-1}},X^{a_{1}},..X^{a_{L}},\psi^{R,j_{H}}],\langle[X^{j_{1}},..X^{j_{H-1}},X^{a_{1}},..X^{a_{L}},
  \psi^{R,j_{H}}]\rangle) - \nonumber \\
  && i^{2(p-N+1)+1}\Sigma_{L=0}^{N} \Sigma_{H=0}^{N-L}\Sigma_{a_{1}..a_{L}=0}^{p}\Sigma_{j_{1}..j_{H}=p+1}^{M}
  (\langle[\gamma^{j_{1}},..\psi^{R,j_{H}},X^{a_{1}},..X^{a_{L}}],[X^{j_{1}},..\psi^{R,j_{H}},X^{a_{1}},..X^{a_{L}}]\rangle)\Big) \label{PP8}
  \end{eqnarray}

 To write the actions in terms of gauge fields and derivatives with respect to fields, we have to use  some laws. Extending the rules in 
 equation (\ref{r3}) for $M$-theory to $N$-dimensional brackets in \textbf{GLNA-theory}, we can obtain the following laws 
 \cite{h2,h3,h4,h5,h10,h11,h12,h13,h14,h15,h16}:

  \begin{eqnarray}
  && \Sigma_{L=0}^{N} \Sigma_{H=0}^{N-L}\Sigma_{a_{1}..a_{L}=0}^{p}\Sigma_{j_{1}..j_{H}=p+1}^{M} \langle[X^{j_{1}},..X^{j_{H-1}},X^{a_{1}},..
  X^{a_{L}},X^{j_{H}}],\langle[X^{j_{1}},..X^{j_{H-1}},X^{a_{1}},..X^{a_{L}},X^{j_{H}}]\rangle=\nonumber \\ &&
   \frac{1}{2}\Sigma_{L=0}^{N} \Sigma_{H=0}^{N-L}\Sigma_{a_{1}..a_{L}=0}^{p}\Sigma_{j_{1}..j_{H}=p+1}^{M}(X^{j_{1}}..X^{j_{H-1}})^{2}
   \langle \partial_{a_{1}}..\partial_{a_{L}}X^{i},\partial_{a_{1}}..\partial_{a_{L}}X^{i}\rangle\nonumber \\
     &&\nonumber \\
       &&\nonumber \\
  && \Sigma_{L=0}^{N} \Sigma_{H=0}^{N-L}\Sigma_{a_{1}..a_{L}=0}^{p}\Sigma_{j_{1}..j_{H}=p+1}^{M}
     \langle[X^{j_{1}},..X^{j_{H}},X^{a_{1}},..X^{a_{L}}],[X^{j_{1}},..X^{j_{H}},X^{a_{1}},..X^{a_{L}}]\rangle=\nonumber \\
         &&
        \Sigma_{L=0}^{N} \Sigma_{H=0}^{N-L}\Sigma_{a_{1}..a_{L}=0}^{p}\Sigma_{j_{1}..j_{H}=p+1}^{M}\frac{\lambda^{2}}{1.2...N}(X^{j_{1}}..X^{j_{H}})^{2}
        \langle F^{a_{1}..a_{L}}, F^{a_{1}..a_{L}}\rangle\nonumber \\
  &&\nonumber \\
    &&\nonumber \\
      &&\nonumber \\
        &&F_{a_{1}..a_{n}}=\partial_{[a_{1}} A_{a_{2}..a_{n}]}=\partial_{a_{1}} A_{a_{2}..a_{n}}-\partial_{a_{2}} A_{a_{1}..a_{n}}+..\nonumber \\
          &&\nonumber \\
            &&\nonumber \\
  &&\Sigma_{m}\rightarrow \frac{1}{(2\pi)^{p}}\int d^{p+1}\sigma \Sigma_{m-p-1}
  i,j=p+1,..,M\quad a,b=0,1,...p\quad m,n=0,..,M~~
 \nonumber \\
      && \nonumber \\
        && \nonumber \\
          && \Sigma_{L=0}^{N} \Sigma_{H=0}^{N-L}\Sigma_{a_{1}..a_{L}=0}^{p}\Sigma_{j_{1}..j_{H}=p+1}^{M} \langle[\gamma^{j_{1}},..X^{j_{H-1}},
          X^{a_{1}},..X^{a_{L}},\psi^{j_{H}}],\langle[X^{j_{1}},..X^{j_{H-1}},X^{a_{1}},..X^{a_{L}},\psi^{j_{H}}]\rangle=\nonumber \\
         &&
     \frac{1}{2}i\Sigma_{L=0}^{N} \Sigma_{H=0}^{N-L}\Sigma_{a_{1}..a_{L}=0}^{p}\Sigma_{j_{1}..j_{H}=p+1}^{M}(X^{j_{1}}..X^{j_{H-1}})^{2}
     \gamma^{a_{L-1}}\langle \partial_{a_{1}}..\partial_{a_{L-1}}\psi^{i},\partial_{a_{1}}..\partial_{a_{L}}\psi^{i}\rangle\nonumber \\
       &&\nonumber \\
         &&\nonumber \\
    && \Sigma_{L=0}^{N} \Sigma_{H=0}^{N-L}\Sigma_{a_{1}..a_{L}=0}^{p}\Sigma_{j_{1}..j_{H}=p+1}^{M}
       \langle[X^{j_{1}},..X^{j_{H}},X^{a_{1}},..X^{a_{L}}],[X^{j_{1}},..X^{j_{H}},X^{a_{1}},..X^{a_{L}}]\rangle=\nonumber \\
           &&
          i\Sigma_{L=0}^{N} \Sigma_{H=0}^{N-L}\Sigma_{a_{1}..a_{L}=0}^{p}\Sigma_{j_{1}..j_{H}=p+1}^{M}\frac{\lambda^{2}}{1.2...N}
          (X^{j_{1}}..X^{j_{H}})^{2}\gamma^{a_{L-1}}\langle \bar{F}^{a_{1}..a_{L-1}}, \bar{F}^{a_{1}..a_{L}}\rangle\nonumber \\
    &&\nonumber \\
      &&\nonumber \\
        &&\nonumber \\
          &&\bar{F}_{a_{1}..a_{n}}=\partial_{[a_{1}} \bar{A}_{a_{2}..a_{n}]}=\partial_{a_{1}} \bar{A}_{a_{2}..a_{n}}-\partial_{a_{2}} 
          \bar{A}_{a_{1}..a_{n}}+..\nonumber \\
            &&\nonumber \\
              &&\nonumber \\
    &&\Sigma_{m}\rightarrow \frac{1}{(2\pi)^{p}}\int d^{p+1}\sigma \Sigma_{m-p-1}
    i,j=p+1,..,M\quad a,b=0,1,...p\quad m,n=0,..,M~~
    \label{P9}
    \end{eqnarray}

    Here $\bar{A}_{a_{2}..a_{n}}$ are fermionic super-partners of gauge bosons $A_{a_{2}..a_{n}}$ and $\psi$  are the fermionic 
    super partners of scalar strings $X$.
     Using the rules of Eq. (\ref{P9}) in action (\ref{PP8} and \ref{P88} ), we derive the following action for $Gp$-branes:

   \begin{eqnarray}
    &&S_{Gp} = -(T_{Gp}) \int dt \sum_{n=1}^{p}\beta_{n}\Big(\delta^{a_{1},a_{2}...a_{n}}_{b_{1}b_{2}....b_{n}}L^{b_{1}}_{a_{1}}...L^{b_{n}}_{a_{n}}
    \Big)^{1/2}\nonumber\\&&
   (L)^{a_{n}}_{b_{n}}= \delta^{a_{n}}_{b_{n}}Tr\Big(\frac{1}{2}i^{2(p-N)}\Sigma_{L=0}^{N} \Sigma_{H=0}^{N-L}\Sigma_{a_{1}..a_{L}=0}^{p}
   \Sigma_{j_{1}..j_{H}=p+1}^{M} (X^{j_{1}}..X^{j_{H-1}})^{2}\langle \partial_{a_{1}}..\partial_{a_{L}}X^{i},\partial_{a_{1}}..\partial_{a_{L}}X^{i}
   \rangle + \nonumber \\ &&
  i^{2(p-N)} \Sigma_{L=0}^{N} \Sigma_{H=0}^{N-L}\Sigma_{a_{1}..a_{L}=0}^{p}\Sigma_{j_{1}..j_{H}=p+1}^{M}\frac{\lambda^{2}}{1.2...N} (X^{j_{1}}..X^{j_{H}})^{2}
  \langle F^{a_{1}..a_{L}}, F^{a_{1}..a_{L}}\rangle  -\nonumber\\ &&
   \frac{1}{2}i^{2(p-N)+1}\Sigma_{L=0}^{N} \Sigma_{H=0}^{N-L}\Sigma_{a_{1}..a_{L}=0}^{p}\Sigma_{j_{1}..j_{H}=p+1}^{M} (X^{j_{1}}..X^{j_{H-1}})^{2}
   \gamma^{a_{L-1}}\langle \partial_{a_{1}}..\partial_{a_{L-1}}\psi^{i},\partial_{a_{1}}..\partial_{a_{L}}\psi^{i}\rangle -\nonumber \\
   &&  i^{2(p-N)+1}\Sigma_{L=0}^{N} \Sigma_{H=0}^{N-L}\Sigma_{a_{1}..a_{L}=0}^{p}\Sigma_{j_{1}..j_{H}=p+1}^{M}\frac{\lambda^{2}}{1.2...N}
   \gamma^{a_{L-1}}(X^{j_{1}}..X^{j_{H}})^{2} \langle \bar{F}^{a_{1}..a_{L-1}}, \bar{F}^{a_{1}..a_{L}}\rangle \Big) \label{P10}
          \end{eqnarray}

   \begin{eqnarray}
   &&S_{Anti-Gp} = -(T_{Anti-Gp}) \int dt \sum_{n=1}^{p}\beta_{n}\Big(
  \delta^{a_{1},a_{2}...a_{n}}_{b_{1}b_{2}....b_{n}}L^{b_{1}}_{a_{1}}...L^{b_{n}}_{a_{n}}\Big)^{1/2}\nonumber\\&&
  (L)^{a_{n}}_{b_{n}}= \delta^{a_{n}}_{b_{n}}Tr\Big(\frac{1}{2}i^{2(p-N+1)}\Sigma_{L=0}^{N} \Sigma_{H=0}^{N-L}\Sigma_{a_{1}..a_{L}=0}^{p}
  \Sigma_{j_{1}..j_{H}=p+1}^{M} (X^{j_{1}}..X^{j_{H-1}})^{2}\langle \partial_{a_{1}}..\partial_{a_{L}}X^{i},\partial_{a_{1}}..\partial_{a_{L}}X^{i}
  \rangle + \nonumber \\ &&
 i^{2(p-N+1)} \Sigma_{L=0}^{N} \Sigma_{H=0}^{N-L}\Sigma_{a_{1}..a_{L}=0}^{p}\Sigma_{j_{1}..j_{H}=p+1}^{M}\frac{\lambda^{2}}{1.2...N} (X^{j_{1}}..X^{j_{H}})^{2}
 \langle \hat{F}^{a_{1}..a_{L}}, \hat{F}^{a_{1}..a_{L}}\rangle  -\nonumber\\ &&
\frac{1}{2}i^{2(p-N+1)+1}\Sigma_{L=0}^{N} \Sigma_{H=0}^{N-L}\Sigma_{a_{1}..a_{L}=0}^{p}\Sigma_{j_{1}..j_{H}=p+1}^{M} (X^{j_{1}}..X^{j_{H-1}})^{2}
\gamma^{a_{L-1}}\langle \partial_{a_{1}}..\partial_{a_{L-1}}\psi^{i},\partial_{a_{1}}..\partial_{a_{L}}\psi^{i}\rangle -\nonumber \\
 &&  i^{2(p-N+1)+1}\Sigma_{L=0}^{N} \Sigma_{H=0}^{N-L}\Sigma_{a_{1}..a_{L}=0}^{p}\Sigma_{j_{1}..j_{H}=p+1}^{M}\frac{\lambda^{2}}{1.2...N}
 \gamma^{a_{L-1}}(X^{j_{1}}..X^{j_{H}})^{2} \langle \hat{\bar{F}}^{a_{1}..a_{L-1}}, \hat{\bar{F}}^{a_{1}..a_{L}}\rangle \Big) \label{PP10}
  \end{eqnarray}

   These actions reduce to the action of $Dp$-branes (\ref{s7}) for $N=2$ and $M=9$ and the action of $Mp$-branes for $N=3$ and $M=10$ 
   \cite{h2,h3,h4,h5,h6,h7,h8,h9,h10,h11,h12,h13,h14,h15,h16}. Also, it is clear that for each scalar, there is a fermion, and 
   for each bosonic gauge field with each rank, there exists a fermionic gauge field with the same rank. This means that these 
   actions are super-symmetric and the number of degrees of freedom for both fermions and bosons are the same. For $\beta_{2}\neq 0$ and 
   $\beta_{n}=0, n\neq 2$, the above actions can be simplified to:

    \begin{eqnarray}
     &&S_{Gp} = -(T_{Gp}) \int dt \beta_{2}\Big(
   \delta^{a_{1},a_{2}}_{b_{1}b_{2}} (\delta^{b_{1}}_{a_{1}}\delta^{b_{2}}_{a_{2}}-\delta^{b_{1}}_{a_{2}}\delta^{b_{2}}_{a_{1}})\times \nonumber \\
    && Tr\Big(\frac{1}{2}i^{2(p-N)}\Sigma_{L=0}^{N} \Sigma_{H=0}^{N-L}\Sigma_{a_{1}..a_{L}=0}^{p}\Sigma_{j_{1}..j_{H}=p+1}^{M} (X^{j_{1}}..X^{j_{H-1}})^{2}
    \langle \partial_{a_{1}}..\partial_{a_{L}}X^{i},\partial_{a_{1}}..\partial_{a_{L}}X^{i}\rangle + \nonumber \\&&
    i^{2(p-N)} \Sigma_{L=0}^{N} \Sigma_{H=0}^{N-L}\Sigma_{a_{1}..a_{L}=0}^{p}\Sigma_{j_{1}..j_{H}=p+1}^{M}
    \frac{\lambda^{2}}{1.2...N} (X^{j_{1}}..X^{j_{H}})^{2}\langle F^{a_{1}..a_{L}}, F^{a_{1}..a_{L}}\rangle  -\nonumber\\ &&
   \frac{1}{2}i^{2(p-N)+1}\Sigma_{L=0}^{N} \Sigma_{H=0}^{N-L}\Sigma_{a_{1}..a_{L}=0}^{p}\Sigma_{j_{1}..j_{H}=p+1}^{M} (X^{j_{1}}..X^{j_{H-1}})^{2}
   \gamma^{a_{L-1}}\langle \partial_{a_{1}}..\partial_{a_{L-1}}\psi^{i},\partial_{a_{1}}..\partial_{a_{L}}\psi^{i}\rangle -\nonumber \\
   &&  i^{2(p-N)+1}\Sigma_{L=0}^{N} \Sigma_{H=0}^{N-L}\Sigma_{a_{1}..a_{L}=0}^{p}\Sigma_{j_{1}..j_{H}=p+1}^{M}\frac{\lambda^{2}}{1.2...N}
   \gamma^{a_{L-1}}(X^{j_{1}}..X^{j_{H}})^{2} \langle \bar{F}^{a_{1}..a_{L-1}}, \bar{F}^{a_{1}..a_{L}}\rangle \Big)\Big) \label{Pr10}
    \end{eqnarray}

   \begin{eqnarray}
   &&S_{Anti-Gp} = -(T_{Anti-Gp}) \int dt \beta_{2}\Big( \delta^{a_{1},a_{2}}_{b_{1}b_{2}} (\delta^{b_{1}}_{a_{1}}\delta^{b_{2}}_{a_{2}}-
   \delta^{b_{1}}_{a_{2}}\delta^{b_{2}}_{a_{1}})\times \nonumber \\
   &&Tr\Big(\frac{1}{2}i^{2(p-N+1)}\Sigma_{L=0}^{N} \Sigma_{H=0}^{N-L}\Sigma_{a_{1}..a_{L}=0}^{p}\Sigma_{j_{1}..j_{H}=p+1}^{M} (X^{j_{1}}..X^{j_{H-1}})^{2}
   \langle \partial_{a_{1}}..\partial_{a_{L}}X^{i},\partial_{a_{1}}..\partial_{a_{L}}X^{i}\rangle + \nonumber \\&&
  i^{2(p-N+1)} \Sigma_{L=0}^{N} \Sigma_{H=0}^{N-L}\Sigma_{a_{1}..a_{L}=0}^{p}\Sigma_{j_{1}..j_{H}=p+1}^{M}\frac{\lambda^{2}}{1.2...N} 
  (X^{j_{1}}..X^{j_{H}})^{2}\langle \hat{F}^{a_{1}..a_{L}}, \hat{F}^{a_{1}..a_{L}}\rangle  -\nonumber\\ &&
  \frac{1}{2}i^{2(p-N+1)+1}\Sigma_{L=0}^{N} \Sigma_{H=0}^{N-L}\Sigma_{a_{1}..a_{L}=0}^{p}\Sigma_{j_{1}..j_{H}=p+1}^{M} (X^{j_{1}}..X^{j_{H-1}})^{2}
  \gamma^{a_{L-1}}\langle \partial_{a_{1}}..\partial_{a_{L-1}}\psi^{i},\partial_{a_{1}}..\partial_{a_{L}}\psi^{i}\rangle -\nonumber \\
 &&  i^{2(p-N+1)+1}\Sigma_{L=0}^{N} \Sigma_{H=0}^{N-L}\Sigma_{a_{1}..a_{L}=0}^{p}\Sigma_{j_{1}..j_{H}=p+1}^{M}\frac{\lambda^{2}}{1.2...N}
 \gamma^{a_{L-1}}(X^{j_{1}}..X^{j_{H}})^{2} \langle \hat{\bar{F}}^{a_{1}..a_{L-1}}, \hat{\bar{F}}^{a_{1}..a_{L}}\rangle \Big)\Big)
  \label{PPr10}
  \end{eqnarray}

 Now, we can extract the wave equations from the actions of (\ref{Pr10}) for branes:

 \begin{eqnarray}
 &&  i^{2(p-N)}\Sigma_{L=0}^{N} \Sigma_{H=0}^{N-L}\Sigma_{a_{1}..a_{L}=0}^{p}\Sigma_{j_{1}..j_{H}=p+1}^{M} (X^{j_{1}}..X^{j_{H-1}})^{2}
 \partial_{a_{1}}^{2}..\partial_{a_{L}}^{2}X^{i} +  \nonumber \\ &&
 i^{2(p-N)}\Sigma_{L=0}^{N} \Sigma_{H=0}^{N-L}\Sigma_{a_{1}..a_{L}=0}^{p}\Sigma_{j_{1}..j_{H}=p+1}^{M} \partial_{a_{1}}..
 \partial_{a_{L}}(X^{j_{1}}..X^{j_{H-1}})^{2}\partial_{a_{1}}..\partial_{a_{L}}X^{i} - \nonumber \\&&
  i^{2(p-N)} \Sigma_{L=0}^{N} \Sigma_{H=0}^{N-L}\Sigma_{a_{1}..a_{L}=0}^{p}\Sigma_{j_{1}..j_{H}=p+1}^{M}
  \frac{\lambda^{2}}{1.2...N} (X^{j_{1}}..X^{j_{H}-1})^{2}\langle F^{a_{1}..a_{L}}, F^{a_{1}..a_{L}}\rangle =0 \nonumber\\ &&\nonumber\\ &&\nonumber\\ &&
 i^{2(p-N)+1}\Sigma_{L=0}^{N} \Sigma_{H=0}^{N-L}\Sigma_{a_{1}..a_{L}=0}^{p}\Sigma_{j_{1}..j_{H}=p+1}^{M} (X^{j_{1}}..X^{j_{H-1}})^{2}
 \gamma^{a_{L-1}} \partial_{a_{1}}^{2}..\partial_{a_{L}-1}^{2}\partial_{a_{L}}\psi^{i}+\nonumber\\ && i^{2(p-N)+1}\Sigma_{L=0}^{N} 
 \Sigma_{H=0}^{N-L}\Sigma_{a_{1}..a_{L}=0}^{p}\Sigma_{j_{1}..j_{H}=p+1}^{M} \partial_{a_{1}}..\partial_{a_{L-1}}(X^{j_{1}}..X^{j_{H-1}})^{2}
 \gamma^{a_{L-1}}\partial_{a_{1}}..\partial_{a_{L}}\psi^{i}-\nonumber \\
 &&  i^{2(p-N)+1}\Sigma_{L=0}^{N} \Sigma_{H=0}^{N-L}\Sigma_{a_{1}..a_{L}=0}^{p}\Sigma_{j_{1}..j_{H}=p+1}^{M}\frac{\lambda^{2}}{1.2...N}
 \gamma^{a_{L-1}}(X^{j_{1}}..X^{j_{H}-1})^{2} \langle \bar{F}^{a_{1}..a_{L-1}}, \bar{F}^{a_{1}..a_{L}}\rangle =0 \label{BV1}
 \end{eqnarray}

  Also, the wave equations for anti-branes can be obtained from the action of (\ref{PPr10}):

  \begin{eqnarray}
  &&  i^{2(p-N)}\Sigma_{L=0}^{N} \Sigma_{H=0}^{N-L}\Sigma_{a_{1}..a_{L}=0}^{p}\Sigma_{j_{1}..j_{H}=p+1}^{M} (X^{j_{1}}..X^{j_{H-1}})^{2}
  \partial_{a_{1}}^{2}..\partial_{a_{L}}^{2}X^{i} +  \nonumber \\ &&
  i^{2(p-N)}\Sigma_{L=0}^{N} \Sigma_{H=0}^{N-L}\Sigma_{a_{1}..a_{L}=0}^{p}\Sigma_{j_{1}..j_{H}=p+1}^{M} \partial_{a_{1}}..
  \partial_{a_{L}}(X^{j_{1}}..X^{j_{H-1}})^{2}\partial_{a_{1}}..\partial_{a_{L}}X^{i} - \nonumber \\&&
  i^{2(p-N)} \Sigma_{L=0}^{N} \Sigma_{H=0}^{N-L}\Sigma_{a_{1}..a_{L}=0}^{p}\Sigma_{j_{1}..j_{H}=p+1}^{M}
  \frac{\lambda^{2}}{1.2...N} (X^{j_{1}}..X^{j_{H}-1})^{2}\langle \hat{F}^{a_{1}..a_{L}}, \hat{F}^{a_{1}..a_{L}}\rangle =0 
  \nonumber\\ &&\nonumber\\ &&\nonumber\\ &&
   i^{2(p-N)+1}\Sigma_{L=0}^{N} \Sigma_{H=0}^{N-L}\Sigma_{a_{1}..a_{L}=0}^{p}\Sigma_{j_{1}..j_{H}=p+1}^{M} (X^{j_{1}}..X^{j_{H-1}})^{2}
   \gamma^{a_{L-1}} \partial_{a_{1}}^{2}..\partial_{a_{L}-1}^{2}\partial_{a_{L}}\psi^{i}+\nonumber\\ && i^{2(p-N)+1}\Sigma_{L=0}^{N} 
   \Sigma_{H=0}^{N-L}\Sigma_{a_{1}..a_{L}=0}^{p}\Sigma_{j_{1}..j_{H}=p+1}^{M} \partial_{a_{1}}..\partial_{a_{L-1}}(X^{j_{1}}..X^{j_{H-1}})^{2}
   \gamma^{a_{L-1}}\partial_{a_{1}}..\partial_{a_{L}}\psi^{i}-\nonumber \\
   &&  i^{2(p-N)+1}\Sigma_{L=0}^{N} \Sigma_{H=0}^{N-L}\Sigma_{a_{1}..a_{L}=0}^{p}\Sigma_{j_{1}..j_{H}=p+1}^{M}\frac{\lambda^{2}}{1.2...N}
   \gamma^{a_{L-1}}(X^{j_{1}}..X^{j_{H}-1})^{2} \langle \hat{\bar{F}}^{a_{1}..a_{L-1}}, \hat{\bar{F}}^{a_{1}..a_{L}}\rangle =0 \label{BV2}
   \end{eqnarray}

 By assuming that all gauge fields are zero ($F=\bar{F}=0$), and substituting $\partial\equiv -iP$ and $X^{i}=e^{ip.x}$, we 
 obtain the following relations from equations (\ref{BV1} and \ref{BV2}):

 \begin{eqnarray}
 &&  \Big(\Sigma_{L=0}^{N} \Sigma_{H=0}^{N-L}\Sigma_{a_{1}..a_{L}=0}^{p}
 \Sigma_{j_{1}..j_{H}=p+1}^{M}(-1)^{a_{1}+..+a_{L}} P_{a_{1},brane,boson}^{2}..P_{a_{L},brane,boson}^{2} +  \nonumber \\ &&
2\Sigma_{L=0}^{N} \Sigma_{H=0}^{N-L}\Sigma_{a_{1}..a_{L}=0}^{p}
\Sigma_{j_{1}..j_{H}=p+1}^{M}(-1)^{a_{1}+..+a_{L}} (((i)^{a_{1}+..+a_{L}}P_{a_{1},brane,boson}..P_{a_{L},brane,boson})^{j_{1}}..\times \nonumber \\ && 
((i)^{a_{1}+..+a_{L}}P_{a_{1},brane,boson}..P_{a_{L},brane,boson})^{j_{H-1}})P_{a_{1},brane,boson}..P_{a_{L},brane,boson}\Big)X^{i} =0 
\nonumber\\ &&\nonumber\\ &&\nonumber\\ &&
 \Big(\Sigma_{L=0}^{N} \Sigma_{H=0}^{N-L}\Sigma_{a_{1}..a_{L}=0}^{p}\Sigma_{j_{1}..j_{H}=p+1}^{M}(-1)^{a_{1}+..+a_{L}}i
 \gamma^{a_{L-1}} P_{a_{1},brane,fermion}^{2}..P_{a_{L}-1,brane,fermion}^{2}P_{a_{L},brane,fermion} +  \nonumber \\ &&
  2\Sigma_{L=0}^{N} \Sigma_{H=0}^{N-L}\Sigma_{a_{1}..a_{L}=0}^{p}\Sigma_{j_{1}..j_{H}=p+1}^{M}(-1)^{a_{1}+..+a_{L}}i
  \gamma^{a_{L-1}} (((i)^{a_{1}+..+a_{L}-1}P_{a_{1},brane,fermion}..P_{a_{L}-1,brane,fermion})^{j_{1}}..\times \nonumber \\ && 
  ((i)^{a_{1}+..+a_{L}-1}P_{a_{1},brane,fermion}..P_{a_{L}-1,brane,fermion})^{j_{H-1}})P_{a_{1},brane,fermion}..P_{a_{L},brane,fermion}\Big)
  \psi^{i} =0\label{BBV1}
 \end{eqnarray}

 \begin{eqnarray}
 &&  \Big(\Sigma_{L=0}^{N} \Sigma_{H=0}^{N-L}\Sigma_{a_{1}..a_{L}=0}^{p}
 \Sigma_{j_{1}..j_{H}=p+1}^{M}(-1)^{a_{1}+..+a_{L}} P_{a_{1},anti-brane,boson}^{2}..P_{a_{L},anti-brane,boson}^{2} +  \nonumber \\ &&
 2\Sigma_{L=0}^{N} \Sigma_{H=0}^{N-L}\Sigma_{a_{1}..a_{L}=0}^{p}
 \Sigma_{j_{1}..j_{H}=p+1}^{M}(-1)^{a_{1}+..+a_{L}} (((i)^{a_{1}+..+a_{L}}P_{a_{1},anti-brane,boson}..P_{a_{L},anti-brane,boson})^{j_{1}}..
 \times \nonumber \\ && 
 ((i)^{a_{1}+..+a_{L}}P_{a_{1},anti-brane,boson}..P_{a_{L},anti-brane,boson})^{j_{H-1}})P_{a_{1},anti-brane,boson}..P_{a_{L},anti-brane,boson}
 \Big)X^{i} =0 \nonumber\\ &&\nonumber\\ &&\nonumber\\ &&
 \Big(\Sigma_{L=0}^{N} \Sigma_{H=0}^{N-L}\Sigma_{a_{1}..a_{L}=0}^{p}\Sigma_{j_{1}..j_{H}=p+1}^{M}(-1)^{a_{1}+..+a_{L}}\times \nonumber\\ && i
 \gamma^{a_{L-1}} P_{a_{1},anti-brane,fermion}^{2}..P_{a_{L}-1,anti-brane,fermion}^{2}P_{a_{L},brane,fermion} +  \nonumber \\ &&
 2\Sigma_{L=0}^{N} \Sigma_{H=0}^{N-L}\Sigma_{a_{1}..a_{L}=0}^{p}\Sigma_{j_{1}..j_{H}=p+1}^{M}(-1)^{a_{1}+..+a_{L}}\times \nonumber\\ && i
 \gamma^{a_{L-1}} (((i)^{a_{1}+..+a_{L}-1}P_{a_{1},anti-brane,fermion}..P_{a_{L}-1,anti-brane,fermion})^{j_{1}}..\times \nonumber \\ &&
 ((i)^{a_{1}+..+a_{L}-1}P_{a_{1},anti-brane,fermion}..P_{a_{L}-1,anti-brane,fermion})^{j_{H-1}})\times \nonumber\\ && P_{a_{1},anti-brane,fermion}..P_{a_{L},anti-brane,fermion}\Big)
 \psi^{i} =0\label{BBV2}
  \end{eqnarray}

  These equations show that momenta for bosons and spinors, and also for branes and anti-branes, are different. For example, if the 
  energy-momentum tensor on the four dimensional brane contains only one energy with extra (i)factor, this tensor on a four 
  dimensional anti-brane includes two components of energy with the extra (i)factor.  Now, we compare these wave equations with 
  wave equations in four dimensions:
                                           
 \begin{eqnarray}
 && \square^{2}X^{i}=0 \Rightarrow P_{0}^{2}X^{i}=0 \nonumber\\&& i\gamma^{\mu}\partial_{\mu}\psi^{i}=0\Rightarrow i\gamma^{\mu}P_{0,\mu}\psi^{i}=0
 \label{BBV3}
\end{eqnarray}

  Comparing the equation (\ref{BBV3}) with equations (\ref{BBV1} and \ref{BBV2}), we can obtain the explicit form of the momenta 
  in a four dimensional universe in terms of the momenta in Lie-$N$-algebra:

 \begin{eqnarray}
 &&  P_{0,Uni, boson}^{2}=\Sigma_{L=0}^{N} \Sigma_{H=0}^{N-L}\Sigma_{a_{1}..a_{L}=0}^{p}
 \Sigma_{j_{1}..j_{H}=p+1}^{M}(-1)^{a_{1}+..+a_{L}} P_{a_{1},brane,boson}^{2}..P_{a_{L},brane,boson}^{2} +  \nonumber \\ &&
 2\Sigma_{L=0}^{N} \Sigma_{H=0}^{N-L}\Sigma_{a_{1}..a_{L}=0}^{p}\Sigma_{j_{1}..j_{H}=p+1}^{M}(-1)^{a_{1}+..+a_{L}} (((i)^{a_{1}+..+a_{L}}
 P_{a_{1},brane,boson}..P_{a_{L},brane,boson})^{j_{1}}..\times \nonumber \\ &&((i)^{a_{1}+..+a_{L}}P_{a_{1},brane,boson}..P_{a_{L},brane,boson})^{j_{H-1}})
 P_{a_{1},brane,boson}..P_{a_{L},brane,boson} \nonumber\\ &&\nonumber\\ &&\nonumber\\ &&
 P_{0,Uni, fermion}= \Sigma_{L=0}^{N} \Sigma_{H=0}^{N-L}\Sigma_{a_{1}..a_{L}=0}^{p}\Sigma_{j_{1}..j_{H}=p+1}^{M}(-1)^{a_{1}+..+a_{L}}\times \nonumber\\ && 
 P_{a_{1},brane,fermion}^{2}..P_{a_{L}-1,brane,fermion}^{2}P_{a_{L},brane,fermion} +  \nonumber \\ &&
 2\Sigma_{L=0}^{N} \Sigma_{H=0}^{N-L}\Sigma_{a_{1}..a_{L}=0}^{p}\Sigma_{j_{1}..j_{H}=p+1}^{M}(-1)^{a_{1}+..+a_{L}} 
 (((i)^{a_{1}+..+a_{L}-1}P_{a_{1},brane,fermion}..P_{a_{L}-1,brane,fermion})^{j_{1}}..\times \nonumber \\ && 
 ((i)^{a_{1}+..+a_{L}-1}P_{a_{1},brane,fermion}..P_{a_{L}-1,brane,fermion})^{j_{H-1}})P_{a_{1},brane,fermion}..P_{a_{L},brane,fermion}\label{BBV4}
 \end{eqnarray}

 \begin{eqnarray}
 && P_{0,anti-Uni, boson}^{2}= \Sigma_{L=0}^{N} \Sigma_{H=0}^{N-L}\Sigma_{a_{1}..a_{L}=0}^{p}
 \Sigma_{j_{1}..j_{H}=p+1}^{M}(-1)^{a_{1}+..+a_{L}} P_{a_{1},anti-brane,boson}^{2}..P_{a_{L},anti-brane,boson}^{2} +  \nonumber \\ &&
 2\Sigma_{L=0}^{N} \Sigma_{H=0}^{N-L}\Sigma_{a_{1}..a_{L}=0}^{p}\Sigma_{j_{1}..j_{H}=p+1}^{M}(-1)^{a_{1}+..+a_{L}} 
 (((i)^{a_{1}+..+a_{L}}P_{a_{1},anti-brane,boson}..P_{a_{L},anti-brane,boson})^{j_{1}}..\times \nonumber \\ && 
 ((i)^{a_{1}+..+a_{L}}P_{a_{1},anti-brane,boson}..P_{a_{L},anti-brane,boson})^{j_{H-1}})P_{a_{1},anti-brane,boson}..P_{a_{L},anti-brane,boson} 
 \nonumber\\ &&\nonumber\\ &&\nonumber\\ &&
  P_{0,anti-Uni, fermion}= \Sigma_{L=0}^{N} \Sigma_{H=0}^{N-L}\Sigma_{a_{1}..a_{L}=0}^{p}\Sigma_{j_{1}..j_{H}=p+1}^{M}(-1)^{a_{1}+..+a_{L}}\times \nonumber \\ && 
  P_{a_{1},anti-brane,fermion}^{2}..P_{a_{L}-1,anti-brane,fermion}^{2}P_{a_{L},brane,fermion} +  \nonumber \\ &&
 2\Sigma_{L=0}^{N} \Sigma_{H=0}^{N-L}\Sigma_{a_{1}..a_{L}=0}^{p}
 \Sigma_{j_{1}..j_{H}=p+1}^{M}(-1)^{a_{1}+..+a_{L}} (((i)^{a_{1}+..+a_{L}-1}\times \nonumber\\ && P_{a_{1},anti-brane,fermion}..P_{a_{L}-1,anti-brane,fermion})^{j_{1}}..
 \times \nonumber \\ && 
 ((i)^{a_{1}+..+a_{L}-1}P_{a_{1},anti-brane,fermion}..P_{a_{L}-1,anti-brane,fermion})^{j_{H-1}})\times \nonumber\\ && P_{a_{1},anti-brane,fermion}..P_{a_{L},anti-brane,fermion}
 \label{BBV5}
 \end{eqnarray}

  These equations show that the  momenta in a four dimensional universe can be given in terms of momenta in Lie-$N$-algebra. The order 
  of terms depends on the number of dimensions of the brane which our universe is placed on,  and also on the dimensions of the algebra 
  of the universe. Also, bosonic momenta are related to lower orders of momenta on the brane with respect to fermionic momenta. On 
  the other hand, energy-momentum tensors on the brane contain less number of timing components (which have extra (i)-factor) with 
  respect to one on the anti-brane. For this reason, the relations for momenta on the branes and anti-branes are different. 
  Using the above relations, the modified commutation relations between position coordinates and momenta are:

  \begin{eqnarray}
   && [x_{0,a_{L}},P_{0,Uni, boson}]= i\hbar \Big(\Sigma_{L=0}^{N} \Sigma_{H=0}^{N-L}\Sigma_{a_{1}..a_{L}=0}^{p}
   \Sigma_{j_{1}..j_{H}=p+1}^{M}(-1)^{a_{1}+..+a_{L}} P_{a_{1},brane,boson}..P_{a_{L}-1,brane,boson} +  \nonumber \\ &&
   2\Sigma_{L=0}^{N} \Sigma_{H=0}^{N-L}\Sigma_{a_{1}..a_{L}=0}^{p}\Sigma_{j_{1}..j_{H}=p+1}^{M}(-1)^{(a_{1}+..+a_{L})/2} 
   (((i)^{(a_{1}+..+a_{L})/2}P_{a_{1},brane,boson}..P_{a_{L},brane,boson})^{j_{1}/2}..\times \nonumber \\ && 
   ((i)^{a_{1}+..+a_{L}}P_{a_{1},brane,boson}..P_{a_{L},brane,boson})^{j_{H-1}/2})P_{a_{1},brane,boson}..P_{a_{L}-1,brane,boson}) 
   \nonumber\\ &&\nonumber\\ &&\nonumber\\ &&
   [x_{0,a_{L}},P_{0,Uni, fermion}]= i\hbar  \Big( \Sigma_{L=0}^{N} \Sigma_{H=0}^{N-L}\Sigma_{a_{1}..a_{L}=0}^{p}
   \Sigma_{j_{1}..j_{H}=p+1}^{M}(-1)^{a_{1}+..+a_{L}}\times \nonumber\\ && 
   P_{a_{1},brane,fermion}^{2}..P_{a_{L}-1,brane,fermion}^{2}P_{a_{L},anti-brane,fermion} +  \nonumber \\ &&
   2\Sigma_{L=0}^{N} \Sigma_{H=0}^{N-L}\Sigma_{a_{1}..a_{L}=0}^{p}\Sigma_{j_{1}..j_{H}=p+1}^{M}(-1)^{a_{1}+..+a_{L}} 
   (((i)^{a_{1}+..+a_{L}-1}P_{a_{1},brane,fermion}..P_{a_{L}-1,brane,fermion})^{j_{1}}..\times \nonumber \\ && 
   ((i)^{a_{1}+..+a_{L}-1}P_{a_{1},brane,fermion}..P_{a_{L}-1,brane,fermion})^{j_{H-1}})P_{a_{1},brane,fermion}..P_{a_{L}-1,brane,fermion}\Big)
   \label{BBV6}
   \end{eqnarray}

   \begin{eqnarray}
   && [x_{0,a_{L}},P_{0,anti-Uni, boson}]= i\hbar \Big(\Sigma_{L=0}^{N} \Sigma_{H=0}^{N-L}\Sigma_{a_{1}..a_{L}=0}^{p}
   \Sigma_{j_{1}..j_{H}=p+1}^{M}(-1)^{a_{1}+..+a_{L}} P_{a_{1},anti-brane,boson}..P_{a_{L}-1,anti-brane,boson} +  \nonumber \\ &&
   2\Sigma_{L=0}^{N} \Sigma_{H=0}^{N-L}\Sigma_{a_{1}..a_{L}=0}^{p}\Sigma_{j_{1}..j_{H}=p+1}^{M}(-1)^{(a_{1}+..+a_{L})/2} 
   (((i)^{(a_{1}+..+a_{L})/2}P_{a_{1},anti-brane,boson}..P_{a_{L},anti-brane,boson})^{j_{1}/2}..\times \nonumber \\ && 
   ((i)^{a_{1}+..+a_{L}}P_{a_{1},anti-brane,boson}..P_{a_{L},anti-brane,boson})^{j_{H-1}/2})P_{a_{1},anti-brane,boson}..P_{a_{L}-1,anti-brane,boson}) 
   \nonumber\\ &&\nonumber\\ &&\nonumber\\ &&
   [x_{0,a_{L}},P_{0,anti-Uni, fermion}]= i\hbar  \Big( \Sigma_{L=0}^{N} \Sigma_{H=0}^{N-L}\Sigma_{a_{1}..a_{L}=0}^{p}
   \Sigma_{j_{1}..j_{H}=p+1}^{M}(-1)^{a_{1}+..+a_{L}}\times \nonumber\\ && 
   P_{a_{1},anti-brane,fermion}^{2}..P_{a_{L}-1,anti-brane,fermion}^{2}P_{a_{L},anti-brane,fermion} +  \nonumber \\ &&
    2\Sigma_{L=0}^{N} \Sigma_{H=0}^{N-L}\Sigma_{a_{1}..a_{L}=0}^{p}\Sigma_{j_{1}..j_{H}=p+1}^{M}(-1)^{a_{1}+..+a_{L}} 
    (((i)^{a_{1}+..+a_{L}-1}P_{a_{1},anti-brane,fermion}..P_{a_{L}-1,anti-brane,fermion})^{j_{1}}..\times \nonumber \\ && 
    ((i)^{a_{1}+..+a_{L}-1}P_{a_{1},anti-brane,fermion}..P_{a_{L}-1,anti-brane,fermion})^{j_{H-1}})P_{a_{1},anti-brane,fermion}..
    P_{a_{L}-1,anti-brane,fermion}\Big)\label{BBV7}
    \end{eqnarray}

  These equations indicate that commutation relations between coordinates and momenta or GUP  in a four dimensional universe can 
  be written in terms of different orders of momenta in a Lie-$N$-algebra. The order of momenta in this version of the GUP is 
  related to the number of dimensions of the brane which our universe is located on. Also, this GUP depends  on the dimensions 
  of the algebra and universe, and has a more exact form with respect to other versions. In this model, bosonic commutation relations depend 
  on lower orders of momenta on the brane with respect to the fermionic one. On the other hand, because of the difference between the number 
  of timing components, the GUP on the brane is different from the one on the anti-brane. Thus, the shape of the GUP depends on the 
  behaviour of particles and changes by changing the properties of the system.

   \section{Calculating the entropy of branes in Lie-N-algebra}\label{o2}
   In this section, we calculate the entropy of branes in the GUP and show that it includes different orders of area of branes 
   from zero to $N^{2}pM$ where N is the dimension of the algebra, p is the dimension of the brane and $M$ is the dimension of 
   the universe. This entropy can be reduced to previous predictions for $N=3$ and eleven dimensions of $M$-theory. To show this, 
   we replace $ \delta p \sim (\frac{\hbar}{\delta x})$ in Eqs. (\ref{BBV4} and \ref{BBV5} ) and obtain:

   \begin{eqnarray}
   &&  E_{0,Uni, boson} \sim \delta P_{0,Uni, boson}=\Big(\Sigma_{L=0}^{N} \Sigma_{H=0}^{N-L}\Sigma_{a_{1}..a_{L}=0}^{p}
   \Sigma_{j_{1}..j_{H}=p+1}^{M}(-1)^{a_{1}+..+a_{L}} (\frac{\hbar}{\delta x})_{a_{1},brane,boson}^{2}..(\frac{\hbar}{\delta x})_{a_{L},brane,boson}^{2} +  
   \nonumber \\ &&
  2\Sigma_{L=0}^{N} \Sigma_{H=0}^{N-L}\Sigma_{a_{1}..a_{L}=0}^{p}\Sigma_{j_{1}..j_{H}=p+1}^{M}(-1)^{a_{1}+..+a_{L}} (((i)^{a_{1}+..+a_{L}}
  (\frac{\hbar}{\delta x})_{a_{1},brane,boson}..(\frac{\hbar}{\delta x})_{a_{L},brane,boson})^{j_{1}}..\times \nonumber \\ && 
  ((i)^{a_{1}+..+a_{L}}(\frac{\hbar}{\delta x})_{a_{1},brane,boson}..(\frac{\hbar}{\delta x})_{a_{L},brane,boson})^{j_{H-1}})
  (\frac{\hbar}{\delta x})_{a_{1},brane,boson}..(\frac{\hbar}{\delta x})_{a_{L},brane,boson}\Big)^{1/2} \nonumber\\ &&\nonumber\\ &&\nonumber\\ &&
  E_{0,Uni, fermion}\sim \delta P_{0,Uni, fermion}= \Sigma_{L=0}^{N} \Sigma_{H=0}^{N-L}\Sigma_{a_{1}..a_{L}=0}^{p}
  \Sigma_{j_{1}..j_{H}=p+1}^{M}(-1)^{a_{1}+..+a_{L}}\times \nonumber\\ && (\frac{\hbar}{\delta x})_{a_{1},brane,fermion}^{2}..
  (\frac{\hbar}{\delta x})_{a_{L}-1,brane,fermion}^{2}(\frac{\hbar}{\delta x})_{a_{L},brane,fermion} +  \nonumber \\ &&
  2\Sigma_{L=0}^{N} \Sigma_{H=0}^{N-L}\Sigma_{a_{1}..a_{L}=0}^{p}\Sigma_{j_{1}..j_{H}=p+1}^{M}(-1)^{a_{1}+..+a_{L}} (((i)^{a_{1}+..+a_{L}-1}
  (\frac{\hbar}{\delta x})_{a_{1},brane,fermion}..(\frac{\hbar}{\delta x})_{a_{L}-1,brane,fermion})^{j_{1}}..\times \nonumber \\ && 
  ((i)^{a_{1}+..+a_{L}-1}(\frac{\hbar}{\delta x})_{a_{1},brane,fermion}..(\frac{\hbar}{\delta x})_{a_{L}-1,brane,fermion})^{j_{H-1}})(\frac{\hbar}
  {\delta x})_{a_{1},brane,fermion}..(\frac{\hbar}{\delta x})_{a_{L},brane,fermion}\label{BBV8}
  \end{eqnarray}

   \begin{eqnarray}
    && E_{0,anti-Uni, boson}\sim\delta P_{0,anti-Uni, boson}= \Big(\Sigma_{L=0}^{N} \Sigma_{H=0}^{N-L}\Sigma_{a_{1}..a_{L}=0}^{p}\times \nonumber \\ && 
    \Sigma_{j_{1}..j_{H}=p+1}^{M}(-1)^{a_{1}+..+a_{L}} (\frac{\hbar}{\delta x})_{a_{1},anti-brane,boson}^{2}..
    (\frac{\hbar}{\delta x})_{a_{L},anti-brane,boson}^{2} +  \nonumber \\ &&
   2\Sigma_{L=0}^{N} \Sigma_{H=0}^{N-L}\Sigma_{a_{1}..a_{L}=0}^{p}\Sigma_{j_{1}..j_{H}=p+1}^{M}(-1)^{a_{1}+..+a_{L}} 
   (((i)^{a_{1}+..+a_{L}}(\frac{\hbar}{\delta x})_{a_{1},anti-brane,boson}..(\frac{\hbar}{\delta x})_{a_{L},anti-brane,boson})^{j_{1}}..
   \times \nonumber \\ &&((i)^{a_{1}+..+a_{L}}(\frac{\hbar}{\delta x})_{a_{1},anti-brane,boson}..
   (\frac{\hbar}{\delta x})_{a_{L},anti-brane,boson})^{j_{H-1}})(\frac{\hbar}{\delta x})_{a_{1},anti-brane,boson}..
   (\frac{\hbar}{\delta x})_{a_{L},anti-brane,boson}\Big)^{1/2} \nonumber\\ &&\nonumber\\ &&\nonumber\\ &&
  E_{0,anti-Uni, fermion} = \delta P_{0,anti-Uni, fermion}= \Sigma_{L=0}^{N} \Sigma_{H=0}^{N-L}\Sigma_{a_{1}..a_{L}=0}^{p}
  \Sigma_{j_{1}..j_{H}=p+1}^{M}(-1)^{a_{1}+..+a_{L}}\times \nonumber \\ && (\frac{\hbar}{\delta x})_{a_{1},anti-brane,fermion}^{2}..
  (\frac{\hbar}{\delta x})_{a_{L}-1,anti-brane,fermion}^{2}(\frac{\hbar}{\delta x})_{a_{L},brane,fermion} +  \nonumber \\ &&
  2\Sigma_{L=0}^{N} \Sigma_{H=0}^{N-L}\Sigma_{a_{1}..a_{L}=0}^{p}\Sigma_{j_{1}..j_{H}=p+1}^{M}(-1)^{a_{1}+..+a_{L}} (((i)^{a_{1}+..+a_{L}-1}
  (\frac{\hbar}{\delta x})_{a_{1},anti-brane,fermion}..(\frac{\hbar}{\delta x})_{a_{L}-1,anti-brane,fermion})^{j_{1}}..\times \nonumber \\ && 
  ((i)^{a_{1}+..+a_{L}-1}(\frac{\hbar}{\delta x})_{a_{1},anti-brane,fermion}..(\frac{\hbar}{\delta x})_{a_{L}-1,anti-brane,fermion})^{j_{H-1}})
  \times\nonumber \\ &&(\frac{\hbar}{\delta x})_{a_{1},anti-brane,fermion}..(\frac{\hbar}{\delta x})_{a_{L},anti-brane,fermion}\label{BBV9}
   \end{eqnarray}

 On the other hand, the area of each brane can be obtained as: ($(\frac{A}{L_{p}^{p}})_{brane}=
 \frac{1}{2}((\frac{A}{L_{p}^{p}})_{brane, fermionic}+(\frac{A}{L_{p}^{p}})_{brane, bosonic})$) 
 in which $(\frac{A}{L_{p}^{p}})_{brane, fermionic}$ denotes the area of the brane which be observed by fermions and 
 $(\frac{A}{L_{p}^{p}})_{brane, bosonic}$ is the area of the brane which is observed by bosons. Also, 
 $(\frac{A}{L_{p}^{p}})_{brane, fermionic/bosonic}\approx \delta x_{1}..\delta x_{p}$ and thus $\delta x\approx (\frac{A}{L_{p}^{p}})^{1/p}$.
 We can rewrite equations (\ref{BBV8} and \ref{BBV9}) in terms of the area of the branes as:

   \begin{eqnarray}
   &&  E_{Uni, boson} \sim \delta P_{0,Uni, boson}=\Big(\Sigma_{L=0}^{N} \Sigma_{H=0}^{N-L}\Sigma_{a_{1}..a_{L}=0}^{p}
   \Sigma_{j_{1}..j_{H}=p+1}^{M}(-1)^{a_{1}+..+a_{L}}\times \nonumber\\&& (\frac{\hbar}{(\frac{A}{L_{p}^{p}})^{1/p}})_{a_{1},brane,boson}^{2}..
   (\frac{\hbar}{(\frac{A}{L_{p}^{p}})^{1/p}})_{a_{L},brane,boson}^{2} +  \nonumber \\ &&
   2\Sigma_{L=0}^{N} \Sigma_{H=0}^{N-L}\Sigma_{a_{1}..a_{L}=0}^{p}\Sigma_{j_{1}..j_{H}=p+1}^{M}(-1)^{a_{1}+..+a_{L}} (((i)^{a_{1}+..+a_{L}}\times \nonumber\\&&
   (\frac{\hbar}{(\frac{A}{L_{p}^{p}})^{1/p}})_{a_{1},brane,boson}..(\frac{\hbar}{(\frac{A}{L_{p}^{p}})^{1/p}})_{a_{L},brane,boson})^{j_{1}}..
   \times \nonumber \\ &&((i)^{a_{1}+..+a_{L}}(\frac{\hbar}{(\frac{A}{L_{p}^{p}})^{1/p}})_{a_{1},brane,boson}..(\frac{\hbar}{(\frac{A}
   {L_{p}^{p}})^{1/p}})_{a_{L},brane,boson})^{j_{H-1}})(\frac{\hbar}{(\frac{A}{L_{p}^{p}})^{1/p}})_{a_{1},brane,boson}..(\frac{\hbar}
   {(\frac{A}{L_{p}^{p}})^{1/p}})_{a_{L},brane,boson}\Big)^{1/2} \nonumber\\ &&\nonumber\\ &&\nonumber\\ &&
   E_{Uni, fermion}\sim \delta P_{0,Uni, fermion}= \Sigma_{L=0}^{N} \Sigma_{H=0}^{N-L}\Sigma_{a_{1}..a_{L}=0}^{p}\Sigma_{j_{1}..j_{H}=p+1}^{M}
   (-1)^{a_{1}+..+a_{L}}\times \nonumber\\ && (\frac{\hbar}{(\frac{A}{L_{p}^{p}})^{1/p}})_{a_{1},brane,fermion}^{2}..(\frac{\hbar}
   {(\frac{A}{L_{p}^{p}})^{1/p}})_{a_{L}-1,brane,fermion}^{2}(\frac{\hbar}{(\frac{A}{L_{p}^{p}})^{1/p}})_{a_{L},brane,fermion} +  \nonumber \\ &&
   2\Sigma_{L=0}^{N} \Sigma_{H=0}^{N-L}\Sigma_{a_{1}..a_{L}=0}^{p}\Sigma_{j_{1}..j_{H}=p+1}^{M}(-1)^{a_{1}+..+a_{L}} (((i)^{a_{1}+..+a_{L}-1}\times \nonumber\\&&
   (\frac{\hbar}{(\frac{A}{L_{p}^{p}})^{1/p}})_{a_{1},brane,fermion}..(\frac{\hbar}{(\frac{A}{L_{p}^{p}})^{1/p}})_{a_{L}-1,brane,fermion})^{j_{1}}..
   \times \nonumber \\ &&((i)^{a_{1}+..+a_{L}-1}(\frac{\hbar}{(\frac{A}{L_{p}^{p}})^{1/p}})_{a_{1},brane,fermion}..(\frac{\hbar}
   {(\frac{A}{L_{p}^{p}})^{1/p}})_{a_{L}-1,brane,fermion})^{j_{H-1}})\times \nonumber\\&&(\frac{\hbar}
   {(\frac{A}{L_{p}^{p}})^{1/p}})_{a_{1},brane,fermion}..(\frac{\hbar}{(\frac{A}{L_{p}^{p}})^{1/p}})_{a_{L},brane,fermion}\nonumber\\&&\label{BBV10}
    \end{eqnarray}

   \begin{eqnarray}
   && E_{anti-Uni, boson}\sim\delta P_{0,anti-Uni, boson}= \Big(\Sigma_{L=0}^{N} \Sigma_{H=0}^{N-L}\Sigma_{a_{1}..a_{L}=0}^{p}\times \nonumber \\ && 
   \Sigma_{j_{1}..j_{H}=p+1}^{M}(-1)^{a_{1}+..+a_{L}} (\frac{\hbar}{(\frac{A}{L_{p}^{p}})^{1/p}})_{a_{1},anti-brane,boson}^{2}..(\frac{\hbar}
   {(\frac{A}{L_{p}^{p}})^{1/p}})_{a_{L},anti-brane,boson}^{2} +  \nonumber \\ &&
   2\Sigma_{L=0}^{N} \Sigma_{H=0}^{N-L}\Sigma_{a_{1}..a_{L}=0}^{p}\Sigma_{j_{1}..j_{H}=p+1}^{M}(-1)^{a_{1}+..+a_{L}} (((i)^{a_{1}+..+a_{L}}
   (\frac{\hbar}{(\frac{A}{L_{p}^{p}})^{1/p}})_{a_{1},anti-brane,boson}..(\frac{\hbar}{(\frac{A}{L_{p}^{p}})^{1/p}})_{a_{L},anti-brane,boson})^{j_{1}}..
   \times \nonumber \\ &&((i)^{a_{1}+..+a_{L}}(\frac{\hbar}{(\frac{A}{L_{p}^{p}})^{1/p}})_{a_{1},anti-brane,boson}..(\frac{\hbar}
   {(\frac{A}{L_{p}^{p}})^{1/p}})_{a_{L},anti-brane,boson})^{j_{H-1}})\times \nonumber\\&&
   (\frac{\hbar}{(\frac{A}{L_{p}^{p}})^{1/p}})_{a_{1},anti-brane,boson}..(\frac{\hbar}{(\frac{A}{L_{p}^{p}})^{1/p}})_{a_{L},anti-brane,boson}\Big)^{1/2} 
   \nonumber\\ &&\nonumber\\ &&\nonumber\\ &&
  E_{anti-Uni, fermion} = \delta P_{0,anti-Uni, fermion}= \Sigma_{L=0}^{N} \Sigma_{H=0}^{N-L}\Sigma_{a_{1}..a_{L}=0}^{p}
  \Sigma_{j_{1}..j_{H}=p+1}^{M}(-1)^{a_{1}+..+a_{L}}\times \nonumber \\ && 
  (\frac{\hbar}{(\frac{A}{L_{p}^{p}})^{1/p}})_{a_{1},anti-brane,fermion}^{2}..(\frac{\hbar}
  {(\frac{A}{L_{p}^{p}})^{1/p}})_{a_{L}-1,anti-brane,fermion}^{2}(\frac{\hbar}{(\frac{A}{L_{p}^{p}})^{1/p}})_{a_{L},brane,fermion} +  \nonumber \\ &&
   2\Sigma_{L=0}^{N} \Sigma_{H=0}^{N-L}\Sigma_{a_{1}..a_{L}=0}^{p}\Sigma_{j_{1}..j_{H}=p+1}^{M}(-1)^{a_{1}+..+a_{L}} (((i)^{a_{1}+..+a_{L}-1}
   (\frac{\hbar}{(\frac{A}{L_{p}^{p}})^{1/p}})_{a_{1},anti-brane,fermion}\times \nonumber \\ &&..(\frac{\hbar}
   {(\frac{A}{L_{p}^{p}})^{1/p}})_{a_{L}-1,anti-brane,fermion})^{j_{1}}..\times \nonumber \\ && 
   ((i)^{a_{1}+..+a_{L}-1}(\frac{\hbar}{(\frac{A}{L_{p}^{p}})^{1/p}})_{a_{1},anti-brane,fermion}..(\frac{\hbar}
   {(\frac{A}{L_{p}^{p}})^{1/p}})_{a_{L}-1,anti-brane,fermion})^{j_{H-1}})\times\nonumber \\ &&(\frac{\hbar}
   {(\frac{A}{L_{p}^{p}})^{1/p}})_{a_{1},anti-brane,fermion}..(\frac{\hbar}{(\frac{A}{L_{p}^{p}})^{1/p}})_{a_{L},anti-brane,fermion}\label{BBV11}                       \end{eqnarray}

   These equations show that the energy of bosons and fermions have a direct relation with different orders of 
   $(\frac{A}{L_{p}^{p}})^{-1}$, where $(\frac{A}{L_{p}^{p}})$ is the area of the universe. This is because for small 
   area, particles are very close to each other, interact more and the energy of the system increases, while for large area, 
   particles become distant, interact less and the energy of the system decreases. Previously it has been shown that 
   the entropy of the system can be obtained from the following relation \cite{h17}:

   \begin{eqnarray}
  && \bar{S}\approx \int d (\frac{A}{L_{p}^{p}}) \frac{(\Delta \bar{S})_{min}}{(\Delta (\frac{A}{L_{p}^{p}}))_{min}} \label{BBV12} 
  \end{eqnarray}

  In agreement with the results of information theory \cite{h17,h18}, we can predict that the minimal increase of entropy should be 
  independent of the value of the area, simply one “bit” of information; let us assume this fundamental unit of entropy as 
  $(\Delta \bar{S})_{min}=b=\ln 2$. On the other hand, for a system which includes particles with energy E and  length 
  $\delta x$, the minimum increase in area can be given by \cite{h17}:
   
   \begin{eqnarray}
    &&  (\Delta (\frac{A}{L_{p}^{p}}))_{min}=8\pi L_{p}^{2}E \delta x=8\pi L_{p}^{2}\frac{1}{2}(E_{Uni, boson}+E_{Uni, fermion}) \delta x = 
    \nonumber\\&& 8\pi L_{p}^{2}\frac{1}{2}(E_{Uni, boson}+E_{Uni, fermion})(\frac{A}{L_{p}^{p}})^{-1/p}  \label{BBV13}
   \end{eqnarray}

   Thus, entropy for branes and anti-branes can be obtained as:

   \begin{eqnarray}
     && \bar{S}_{Universe}\approx \int d (\frac{A}{L_{p}^{p}}) \frac{b}{8\pi L_{p}^{2}\frac{1}{2}(E_{Uni, boson}+E_{Uni, fermion})
     (\frac{A}{L_{p}^{p}})^{-1/p}}\approx \nonumber\\&&
   \frac{b}{8\pi L_{p}^{2}} \Big(\Sigma_{L=0}^{N} \Sigma_{H=0}^{N-L}\Sigma_{a_{1}..a_{L}=0}^{p}\Sigma_{j_{1}..j_{H}=p+1}^{M}(-1)^{(a_{1}+..+a_{L})/2} 
   (\frac{1}{(a_{1}+..a_{L})/p+1})(\hbar^{-(a_{1}+..a_{L})}(\frac{A}{L_{p}^{p}})_{brane,boson}^{(a_{1}+..a_{L})/p+1}) +  \nonumber\\&&
   2\Sigma_{L=0}^{N} \Sigma_{H=0}^{N-L}\Sigma_{a_{1}..a_{L}=0}^{p}\Sigma_{j_{1}..j_{H}=p+1}^{M}(-1)^{(a_{1}+..+a_{L})/2} (((i)^{(a_{1}+..a_{L})
   (j_{1}+...j_{H-1}+1)/2}(\hbar^{(a_{1}+..a_{L})(j_{1}+...j_{H-1}+1)}\times \nonumber\\&&(\frac{1}{(a_{1}+..a_{L})(j_{1}+...j_{H-1}+1)/2p+1}) 
   (\frac{A}{L_{p}^{p}})_{brane,boson}^{(a_{1}+..a_{L})(j_{1}+...j_{H-1}+1)/2p+1}\Big) +\nonumber\\&&\Big(\Sigma_{L=0}^{N} \Sigma_{H=0}^{N-L}
   \Sigma_{a_{1}..a_{L}=0}^{p}\Sigma_{j_{1}..j_{H}=p+1}^{M}(-1)^{(a_{1}+..+a_{L})}\times \nonumber\\&& (\frac{1}{(2a_{1}+..2a_{L-1}+a_{L})/p+1})
   (\hbar^{-(2a_{1}+..2a_{L-1}+a_{L})}(\frac{A}{L_{p}^{p}})_{brane,fermion}^{(2a_{1}+..2a_{L-1}+a_{L})/p+1}) +  \nonumber\\&&
   2\Sigma_{L=0}^{N} \Sigma_{H=0}^{N-L}\Sigma_{a_{1}..a_{L}=0}^{p}\Sigma_{j_{1}..j_{H}=p+1}^{M}(-1)^{(a_{1}+..+a_{L})/2} 
   (((i)^{(a_{1}+..a_{L-1})(j_{1}+...j_{H-1}+1)}(\hbar^{(a_{1}+..a_{L})(j_{1}+...j_{H-1}+1)}\times \nonumber\\&& 
   (\frac{1}{((a_{1}+..a_{L-1})(j_{1}+...j_{H-1}+1)+a_{L})/p+1}) 
   (\frac{A}{L_{p}^{p}})_{brane,fermion}^{((a_{1}+..a_{L-1})(j_{1}+...j_{H-1}+1)+a_{L})/p+1}\Big) \label{BBV14} 
   \end{eqnarray}
 
 \begin{eqnarray}
     && \bar{S}_{anti-Universe}\approx \int d (\frac{A}{L_{p}^{p}}) \frac{b}{8\pi L_{p}^{2}\frac{1}{2}(E_{anti-Uni, boson}+E_{anti-Uni, fermion})
     (\frac{A}{L_{p}^{p}})^{-1/p}}\approx \nonumber\\&&
   \frac{b}{8\pi L_{p}^{2}} \Big(\Sigma_{L=0}^{N} \Sigma_{H=0}^{N-L}\Sigma_{a_{1}..a_{L}=0}^{p}\Sigma_{j_{1}..j_{H}=p+1}^{M}(-1)^{(a_{1}+..+a_{L})/2} 
   (\frac{1}{(a_{1}+..a_{L})/p+1})(\hbar^{-(a_{1}+..a_{L})}(\frac{A}{L_{p}^{p}})_{anti-brane,boson}^{(a_{1}+..a_{L})/p+1}) +  \nonumber\\&&
   2\Sigma_{L=0}^{N} \Sigma_{H=0}^{N-L}\Sigma_{a_{1}..a_{L}=0}^{p}\Sigma_{j_{1}..j_{H}=p+1}^{M}(-1)^{(a_{1}+..+a_{L})/2} 
   (((i)^{(a_{1}+..a_{L})(j_{1}+...j_{H-1}+1)/2}(\hbar^{(a_{1}+..a_{L})(j_{1}+...j_{H-1}+1)}\times \nonumber\\&& 
   (\frac{1}{(a_{1}+..a_{L})(j_{1}+...j_{H-1}+1)/2p+1}) (\frac{A}{L_{p}^{p}})_{anti-brane,boson}^{(a_{1}+..a_{L})(j_{1}+...j_{H-1}+1)/2p+1}\Big) + 
   \nonumber\\&&\Big(\Sigma_{L=0}^{N} \Sigma_{H=0}^{N-L}\Sigma_{a_{1}..a_{L}=0}^{p}\Sigma_{j_{1}..j_{H}=p+1}^{M}\times \nonumber\\&& 
   (-1)^{(a_{1}+..+a_{L})} (\frac{1}{(2a_{1}+..2a_{L-1}+a_{L})/p+1})(\hbar^{-(2a_{1}+..2a_{L-1}+a_{L})}
   (\frac{A}{L_{p}^{p}})_{anti-brane,fermion}^{(2a_{1}+..2a_{L-1}+a_{L})/p+1}) +  \nonumber\\&&
   2\Sigma_{L=0}^{N} \Sigma_{H=0}^{N-L}\Sigma_{a_{1}..a_{L}=0}^{p}\Sigma_{j_{1}..j_{H}=p+1}^{M}(-1)^{(a_{1}+..+a_{L})/2} 
   (((i)^{(a_{1}+..a_{L-1})(j_{1}+...j_{H-1}+1)}(\hbar^{(a_{1}+..a_{L})(j_{1}+...j_{H-1}+1)}\times \nonumber\\&& 
   (\frac{1}{((a_{1}+..a_{L-1})(j_{1}+...j_{H-1}+1)+a_{L})/p+1}) 
   (\frac{A}{L_{p}^{p}})_{anti-brane,fermion}^{((a_{1}+..a_{L-1})(j_{1}+...j_{H-1}+1)+a_{L})/p+1}\Big) \label{BBV15}                                                             
   \end{eqnarray}

   These equations show that the predicted entropy on the universe is different from the one on the anti-universe. This is 
   because  areas of the brane on which our universe is located   have less timing coordinates with respect to areas of the 
   anti-brane on which the anti-universe is placed. On the other hand, the surface area, which is seen by fermions is different 
   from the surface that is observed by bosons. These entropies depend on fermionic and bosonic areas, whose order changes 
   from zero to higher values depending on the number of dimensions of branes and the Lie-algebra.

\section{Summary and conclusion }\label{o4}
In this research,  we have shown that the GUP can arise during the formation of branes and anti-branes  in an $M$-dimensional 
universe whose  fields obey a Lie-$N$-algebra. The order of terms in the GUP  changes from zero to higher numbers whose  
values depend on the number of dimensions of branes ($p$), number of dimensions of the universe ($M$) and dimensions of the 
Lie-algebra ($N$). For $N=3$ and an $11$-dimensional universe, the results are consistent with $M$-theory. In this model, 
firstly, two energies with opposite signs are created from nothing, excited and produce various branes and anti-branes with 
different timing dimensions and quantum numbers. The related actions for these branes contain various derivatives of bosonic 
fields which cause the production of terms with different orders of momenta in momentum space. This is known as the generalized 
uncertainty principle or GUP. By compacting branes, various derivatives of fermions are produced which lead to the creation of 
different orders of momenta for spinors, and consequently the emergence of a fermionic GUP. We have calculated the entropy of the 
branes, and find that the number of  terms depend upon the the dimension of the branes and the applied algebra. In fact, with 
increasing the number of dimensions of the branes, higher orders of their areas appear, and the entropy increases.

\section*{Acknowledgments}
\noindent The work of Alireza Sepehri has been supported
financially by the Research Institute for Astronomy and Astrophysics
of Maragha (RIAAM), Iran under research project No.1/4717-97. 
F. Rahaman and A. Pradhan are grateful to IUCAA, Pune, India 
for awarding Visiting Associateship. The work was partially supported 
by VEGA grant No. 2/0009/16 for which R. Pincak thanks the TH division in CERN.
A. Pradhan would also like to thank the University of Zululand, Kwa-Dlangezwa 3886, 
South Africa for providing facilities and support where part of this work has been done.


\begin{thebibliography}{99}
\bibitem{g1}D. Amati, M. Ciafaloni and G. Veneziano, Can spacetime be probed below the string size? {\it Phys. Lett.
B} {\bf 216} (1989) 41; M. Maggiore, Generalized uncertainty principle in quantum gravity, {\it Phys. Lett. B} {\bf 304}
(1993) 65 [hep-th/9301067]; M. Maggiore, Quantum groups, gravity, and the generalized uncertainty principle, 
{\it Phys. Rev. D} {\bf 49} (1994) 5182 [hep-th/9305163]; M. Maggiore, The algebraic structure of the generalized uncertainty principle, 
{\it Phys. Lett. B} {\bf 319} (1993) 83 [hep-th/9309034]; L. J. Garay,Quantum gravity and minimum length
{\it Int. J. Mod. Phys. A} {\bf 10} (1995) 145 [gr-qc/9403008];
F. Scardigli, Generalized uncertainty principle in quantum gravity from micro-black hole gedanken experiment, {\it Phys. Lett. B} 
{\bf 452} (1999) 39 [hep-th/9904025]; 
S. Hossenfelder, M. Bleicher, S. Hofmann, J. Ruppert, S. Scherer and H. Stoecker, Signatures in the planck regime, {\it Phys. Lett. B} 
{\bf 575} (2003) 85 [hep-th/0305262]; 
C. Bambi and F. R. Urban, Natural extension of the generalised uncertainty principle, {\it Class. Quant. Grav.} {\bf 25} (2008) 095006 [gr-qc/0709.1965]; 
G. M. Hossain, V. Husain and S. S. Seahra, Background independent quantization and the uncertainty principle, {\it Class. Quant. Grav.} 
{\bf 27} (2010) 165013 [arXiv:1003.2207 [gr-qc]].
\bibitem{g2}A. Kempf, Nonpointlike particles in harmonic oscillators, {\it J. Phys. A} {\bf 30} (1997) 2093 [hep-th/9604045].
\bibitem{g3}A. Kempf, G. Mangano and R. B. Mann, Hilbert space representation of the minimal length uncertainty relation, {\it Phys. Rev. D} {\bf 52} 
(1995) 1108 [hep-th/9412167].
\bibitem{g4}F. Brau, Minimal length uncertainty relation and hydrogen atom, {\it J. Phys. A} {\bf 32} (1999) 7691 [quant-ph/9905033].
\bibitem{g5}J. Magueijo and L. Smolin, Lorentz invariance with an invariant energy scale, {\it Phys. Rev. Lett.} {\bf 88} (2002) 190403 [hep-th/0112090]; 
J. Magueijo and L. Smolin, String theories with deformed energy momentum relations, and a possible non-tachyonic bosonic string, {\it Phys. Rev. D} {\bf 71} 
(2005) 026010 [hep-th/0401087].
\bibitem{g6}J. L. Cortes and J. Gamboa, Quantum uncertainty in doubly special relativity, {\it Phys. Rev. D} {\bf 71} (2005) 065015 [hep-th/0405285].
\bibitem{g7}S. Ghosh and P. Pal, Deformed special relativity and deformed symmetries in a canonical framework, {\it Phys. Rev. D} {\bf 75} (2007) 
105021 [hep-th/0702159].
\bibitem{g8}S. Das and E. C. Vagenas, Universality of quantum gravity corrections, {\it Phys. Rev. Lett.} {\bf 101} (2008) 
221301 [hep-th/0810.5333].
\bibitem{g9}S. Das and E. C. Vagenas, Phenomenological implications of the generalized uncertainty principle, {\it Can. J. Phys.} {\bf 87} 
(2009) 233 [hep-th/0901.1768].
\bibitem{g10}A. F. Ali, S. Das and E. C. Vagenas, Discreteness of space from the generalized uncertainty principle, {\it Phys. Lett. B} {\bf 678} 
(2009) 497 [hep-th/0906.5396].
\bibitem{g11}S. Das, E. C. Vagenas and A. F. Ali, Discreteness of space from GUP II: relativistic wave equations, {\it Phys. Lett. B} {\bf 690} (2010) 
407 [hep-th/1005.3368].
\bibitem{g12}A. F. Ali, S. Das and E. C. Vagenas, The generalized uncertainty principle and quantum gravity phenomenology, [hep-th/1001.2642].
\bibitem{g13}A. F. Ali, S. Das and E. C. Vagenas, A proposal for testing quantum gravity in the lab, {\it Phys. Rev. D} {\bf 84} (2011) 044013 [hep-th/1107.3164].
\bibitem{g14}Wissam Chemissany, Saurya Das, Ahmed Farag Ali and Elias C. Vagenas, Effect of the generalized uncertainty principle on post-inflation 
preheating, {\it Jour. Cosmology Astropart. Phys.} {\bf 1112} (2011) 017.
\bibitem{h2}Alireza Sepehri, Cosmology from quantum potential in brane-anti-brane system, {\it Phys. Lett B} {\bf 748} (2015) 328335, [gr-qc/1508.01407].
\bibitem{h3}Alireza Sepehri, Born–Infeld extension of Lovelock brane gravity in the system of M0-branes and its application for the emergence of Pauli 
exclusion principle in BIonic superconductors, {\it Phys. Lett. A} {\bf 38} (2016) 2247. 
\bibitem{h4}A. Sepehri, F. Rahaman, S. Capozziello, A. Farag Ali and A. Pradhan, Emergence and oscillation of cosmic space by 
joining M1-branes, {\it Eur. Phys. J. C} {\bf 76} (2016) 231 [gr-qc/1604.0245].
\bibitem{h5}A. Sepehri, M. R. Setare and S. Capozziello, Emergence and expansion of cosmic space as due to M0-branes, 
{\it Eur. Phys. J. C} {\bf 75} (2015) 618 [hep-th/1512.04840].
\bibitem{h6}A. Sepehri, A. Pradhan, A. Beesham and Jaume de Haro, Teleparallel loop quantum cosmology in a system of intersecting branes, 
{\it Phys. Lett. B} {\bf 760} (2016) 94 [gr-qc/1605.02590].
\bibitem{h7}A. Sepehri, F. Rahaman, A. Pradhan, S. Capozziello and I. H. Sardar, {\it Phys. Lett. B} 747 (2015) 1 [gr-qc/15505.05105].
 \bibitem{h8} A. Sepehri, F. Rahaman, A. Pradhan and I. H. Sardar, Emergence and expansion of cosmic space in BIonic system, 
 {\it Phys. Lett. B} 741, 92 (2015) [gr-qc/1505.00428].
\bibitem{h9}A. Sepehri and R. Pincak,The birth of the universe in a new  G-theory  approach, {\it Mod. Phys. Lett. A} {\bf 32} (2017) 1750033 [arXiv:1610.09277].
\bibitem{h10} R. C. Myers, Dielectric-Branes, {\it J. High Energy Phys.} {\bf 9912} (1999) 022 [hep-th/9910053].
\bibitem{h11} N. R. Constable, R. C. Myers and O. Tafjord, Non-abelian brane intersection, {\it J. High Energy Phys.} {\bf 06} (2001) 023; 
A. A. Tseytlin, Born-Infeld action, supersymmetry and string theory, preprint (1999) [hep-th/9908105].
\bibitem{h12}C.-S. Chu and D. J. Smith, Towards the quantum geometry of the M5-brane in a constant $C$-field from multiple membranes, 
{\it J. High Energy Phys.} {\bf 0904} (2009) 097.
\bibitem{h13} J. Bagger and N. Lambert, Gauge Symmetry and Supersymmetry of Multiple M2-Branes,
{\it Phys. Rev. D} {\bf 77} (2008) 065008 [hep-th/0711.0955].
\bibitem{h14} A. Gustavsson, Algebraic structures on parallel $M2$-branes, arXiv:0709.1260 [hep-th]; 
A. Sepehri, R. Pincak, Emergence of the world with Lie-N-algebra and M-dimensions from nothing, arXiv:1610.09257.
\bibitem{h15} Pei-Ming Ho and Yutaka Matsuo, $M5$ from $M2$, {\it J. High Energy Phys.} {\bf 0806} (2008) 105.
\bibitem{h16}S. Mukhi and C. Papageorgakis, $M2$ to $D2$, {\it J. High Energy Phys.} {\bf 05} (2008) 085.
\bibitem{h17}A. J. M. Medved and Elias C. Vagenas, When conceptual worlds collide: The generalized uncertainty principle and the 
Bekenstein-Hawking entrpy, {\it Phys. Rev. D} {\bf 70} (2004) 124021.
\bibitem{h18}C. Adami, The physics of information, preprint (2004) [quant-ph/0405005].

\end{thebibliography}
\end{document}